\documentclass[11pt]{article}
\usepackage{todonotes}
\usepackage[T1]{fontenc}
\usepackage{graphicx}
\usepackage{subcaption}
\usepackage{mathtools}
\usepackage{soul,comment,cite}
\usepackage{amsthm}
\usepackage{amsmath}
\usepackage{amssymb}
\usepackage{systeme}
\usepackage{gensymb}
\usepackage[ruled,vlined]{algorithm2e}
\usepackage{fullpage}
\usepackage[absolute]{textpos}
\usepackage[top=0.8in, bottom=0.9in, left=0.9in, right=0.9in]{geometry}

\newtheorem{theorem}{Theorem}

\newtheorem{lemma}{Lemma}
\newtheorem{observation}{Observation}

\newtheorem{definition}{Definition}
\newtheorem{prop}{Proposition}

\graphicspath{{./figures/}}

\usepackage[pdftex, plainpages = false, pdfpagelabels,
                 bookmarks=false,
                 bookmarksopen = true,
                 bookmarksnumbered = true,
                 breaklinks = true,
                 linktocpage,
                 pagebackref,
                 colorlinks = true,
                 linkcolor = blue,
                 urlcolor  = cyan,
                 citecolor = red,
                 anchorcolor = green,
                 hyperindex = true,
                 hyperfigures
                 ]{hyperref}

\def\bbR{\mathbb{R}}

\def\cd{\mathcal{D}}
\def\ch{\mathcal{H}}
\def\ct{\mathcal{T}}
\def\cp{\mathcal{P}}
\def\calR{\mathcal{R}}

\newcommand{\ol}[1]{\overline{#1}}

\title{Computing Independent Sets of Size Three in Disk Graphs\thanks{A preliminary version of this paper will appear in {\em Proceedings of the 21st International Conference and Workshops on Algorithms and Computation (WALCOM 2027).}}}

\author{
Anastasiia Tkachenko\thanks{Kahlert School of Computing,
University of Utah, Salt Lake City, UT 84112, USA. {\tt anastasiia.tkachenko@utah.edu}}
\and
Haitao Wang\thanks{Kahlert School of Computing,
University of Utah, Salt Lake City, UT 84112, USA. {\tt haitao.wang@utah.edu}}
}

\date{}

\begin{document}

\maketitle

\vspace{-0.2in}

\begin{abstract}
We study the problem of computing an independent set of size three in disk graphs, where vertices correspond to disks in the plane and two vertices are adjacent if their disks intersect. While computing a maximum independent set in disk graphs is NP-hard, the fixed-size case admits more efficient solutions. Given a set of $n$ disks, we present an algorithm to compute an independent set of size three in the disk graph and our algorithm runs in randomized expected $O^*(n^{5/3})$ time, where $O^*$ hides a factor of $n^{\epsilon}$ for an arbitrarily small $\epsilon > 0$. Our approach introduces a framework based on semi-algebraic range searching and tree-structured biclique partitions, which allows us to group pairs of disjoint disks into canonical subsets and avoid redundant computations.

When the centers of the disks are in convex position, we solve the problem in $O(n \log^2 n)$ deterministic time. We also consider the case in which the disks themselves are in convex position, i.e., each disk contributes to the boundary of the convex hull of all disks, and obtain an $O(n \log^2 n)$ time algorithm for this setting as well. Our techniques for these cases are based on geometric properties of disk arrangements. In particular, they rely on new structural lemmas that reduce the problem to efficiently testing certain ordered configurations of disks.

Our results improve upon the straightforward quadratic time approach and demonstrate how geometric structures and range searching techniques can be leveraged to solve fixed-size independent set problems in disk graphs more efficiently. To the best of our knowledge, this is the first systematic study of independent sets of size three in disk graphs. Our results show that even this smallest nontrivial case exhibits rich structural properties, especially under convexity assumptions.
\end{abstract}


\section{Introduction}
\label{sec:intro}

A \emph{geometric intersection graph} is a graph in which vertices represent geometric objects and two vertices are adjacent if the corresponding objects intersect. Geometric intersection graphs form a broad class that captures many natural graph families arising from segments, boxes, circular arcs, disks, balls, and other objects.

A large body of work on such graphs has focused on fundamental global optimization problems, including maximum independent set, maximum clique, and vertex cover. These problems are central from both theoretical and practical perspectives, yet they are NP-hard on general graphs~\cite{ref:KarpRe72}. This hardness persists for several geometric intersection graph classes. For example, the maximum independent set problem remains NP-complete on unit-disk graphs~\cite{ref:ClarkUn90}. Consequently, many approximation algorithms have been developed; see, for example,~\cite{ref:DasDi20,ref:DasAp15,ref:MaratheSi95,ref:MatsuiAp98,ref:ChanAp12,ref:AgarwalIn06}.

On the other hand, computational geometry has a long tradition of developing exact subquadratic, and often near-linear, algorithms for finding a \emph{pair} of geometric objects satisfying a desired relation~\cite{ref:AgarwalTh98, ref:ChezelleDi93, ref:SharirA97,ref:AgarwalAn08,ref:KamousiCl11,ref:SharirIn85, ref:WangIm22, ref:ChoOp24Journal, ref:WangOn22,ref:AgarwalEu91,ref:AgarwalTh98}. A representative example is the \emph{diametral-pair problem}: given $n$ points in the plane, find two points whose distance is maximum. This is exactly the $k=2$ case of the dispersion (or maximally separated set) problem~\cite{ref:AgarwalCo06,ref:SingireddyAl23,ref:TkachenkoDo25}: Given an integer $k$, compute a subset of $k$ points that maximizes the minimum pairwise distance.

But what can be said about problems involving \emph{triples} of geometric objects that must simultaneously satisfy certain pairwise conditions? Continuing the above example, the natural size-$3$ analogue of the diametral-pair problem asks for three objects whose pairwise relations are all satisfied, for example, three points maximizing the minimum pairwise distance or three disks that are pairwise disjoint.

This line of work is motivated by two observations. First, small substructures often capture meaningful local geometric behavior~\cite{ref:ChanFi23}. Second, hardness results for the corresponding global optimization problems typically do not shed light on the complexity of fixed-size cases. For example, the NP-hardness of maximum independent set does not explain the complexity of deciding whether an independent set of size $3$ exists. Consequently, constant-size problems can exhibit structural and algorithmic behavior that is remarkably different from that of their unrestricted counterparts.

Recently, constant-size problems in geometric intersection graphs have received increasing attention. In SODA'23, Chan initiated a systematic study of this direction~\cite{ref:ChanFi23}. For instance, he gave subquadratic algorithms for finding a $3$-cycle, an independent set of size $3$, and a $4$-clique in the intersection graph of $n$ line segments in the plane. In a complementary fine-grained direction, Chan, He, and Yu~\cite{ref:ChanOn23} studied small-size geometric set cover and discrete $k$-center, showing that even very small constants admit nontrivial upper and lower bounds.


\paragraph{Our problem.}
In this paper, we follow this direction and study the problem of finding an independent set of size $3$ in \emph{disk graphs}. A disk graph is the intersection graph of a finite family of disks in the plane. Disk graphs form an important geometric graph class due to their simple yet expressive structure, and they have found numerous applications in areas such as wireless and sensor networks~\cite{ref:ClarkUn90,ref:PerkinsHi94,ref:PerkinsAd99,ref:BalisterCo05}. By allowing arbitrary radii, disk graphs strictly generalize {\em unit-disk graphs} (i.e., graphs in which all disks have the same radius).

The size-$3$ independent set problem is interesting in its own right. In disk graphs, it asks whether a given set of disks contains three pairwise disjoint disks. This is the first genuinely nontrivial fixed-size independent-set problem in this setting. Indeed, size $1$ is trivial, while size $2$ asks only whether there exists a disjoint pair, which can be solved in $O(n \log n)$ time, for example using farthest-disk Voronoi diagrams~\cite{ref:RappaportCo89}. In contrast, size $3$ is the first case in which mutual compatibility among multiple objects must be enforced simultaneously.

\begin{figure}[t]
\begin{minipage}[h]{0.49\textwidth}
\begin{center}
\includegraphics[height=1.8in]{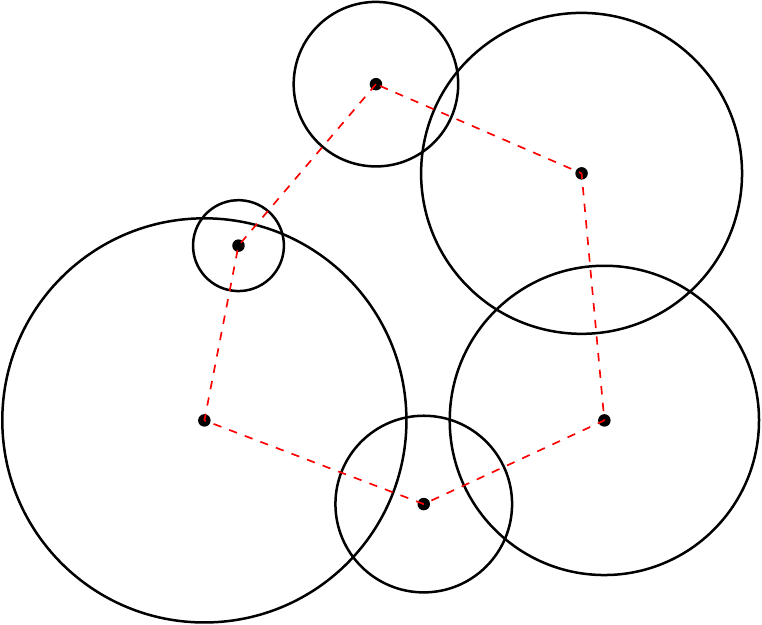}
\caption{\footnotesize A set of disks in center-convex position.}
\label{fig:centerconvex}
\end{center}
\end{minipage}
\hspace{0.05in}
\begin{minipage}[h]{0.49\textwidth}
\begin{center}
\includegraphics[height=1.8in]{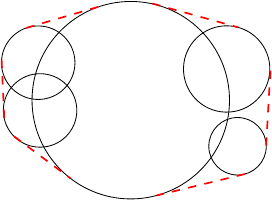}
\caption{\footnotesize A set of disks in disk-convex position.}
\label{fig:convexpos}
\end{center}
\end{minipage}
\vspace{-0.1in}
\end{figure}

We also consider two special settings of the problem: the \emph{center-convex case} and the \emph{disk-convex case}. In the center-convex case, the centers of all disks are in convex position, that is, each center appears as a vertex of the convex hull of the centers (see Figure~\ref{fig:centerconvex}). In the disk-convex case, the disks themselves are in convex position, meaning that each disk contributes to the boundary of the convex hull of all disks (see Figure~\ref{fig:convexpos}).

\paragraph{Related work.}
Independent sets of small size have been studied for several geometric intersection graph classes, including segment intersection graphs and axis-aligned box intersection graphs~\cite{ref:ChanFi23}, as well as unit-disk graphs. For unit-disk graphs, an $O(n^2)$-time algorithm for computing an independent set of size $3$ was given in~\cite{ref:KobayashiAn22}. Agarwal, Overmars, and Sharir presented an $O(n^{4/3}\log^2 n)$-time algorithm~\cite{ref:AgarwalCo06}. Their approach relies on a grid-based technique, which does not extend to the more general setting of disks with arbitrary radii.

When all disks are unit disks and their centers are in convex position, Tkachenko and Wang gave an $O(n \log n)$-time algorithm for computing an independent set of size $3$~\cite{ref:TkachenkoDo25}. Note that in the unit-disk setting, the center-convex and disk-convex cases coincide.

There has been growing interest in studying geometric problems under the \emph{convex-position assumption}, a well-known configuration in computational geometry. In the context of disk and unit disk graphs, this configuration has been explored for several classical problems that are NP-hard in the general case. Under convex-position assumptions, polynomial-time algorithms have been obtained for problems such as the $k$-center problem for points in convex position~\cite{ref:ChoiEf23,ref:TkachenkoDo25}, the maximum independent set problem for unit disks with centers in convex position~\cite{ref:TkachenkoDo25} and for disks in the disk-convex case~\cite{ref:TkachenkoMa26}, the dominating set problem for unit disks with centers in convex position~\cite{ref:TkachenkoDo25} and for disks in the disk-convex case~\cite{ref:TkachenkoCo26}, and the dispersion problem for points in convex position~\cite{ref:SingireddyAl23,ref:TkachenkoDo25}.

The convex-position assumption has also been studied for problems that are already polynomial-time solvable in general, such as computing Voronoi diagrams of point sets~\cite{ref:AggarwalA89}, finding maximum cliques in unit-disk graphs~\cite{ref:TkachenkoCl26}, computing triangulations~\cite{ref:LingasOn86}, and other related problems~\cite{ref:RichardsA90,ref:ChazelleIm93,ref:HerreraIn21,ref:HerreraIn21,ref:HuemerTh20}.

Despite these progress, disk graphs with \emph{arbitrary radii} remain much less understood from the perspective of constant-size independent sets. While related questions have been studied extensively for unit disk graphs, to the best of our knowledge no analogous results were previously known for computing independent sets of size $3$ in general disk graphs. Our work aims to fill this gap.

\paragraph{Our contributions.}
We study the problem of finding an independent set of size $3$ in disk graphs. We propose a randomized algorithm that computes such a set, if one exists, in $O^*(n^{5/3})$ expected time. The algorithm is based on a new framework that combines semi-algebraic range searching~\cite{ref:AgarwalIn25} with tree-structured biclique partitions. This framework allows us to group disjoint pairs of disks into canonical subsets and avoid the redundant computations inherent in naive pairwise enumeration.

We also obtain faster algorithms under natural convexity assumptions. In the center-convex case, where the disk centers are in convex position, we give an $O(n \log^2 n)$-time deterministic algorithm. In the disk-convex case, where each disk contributes to the boundary of the convex hull of all disks, we obtain the same time bound. These running times are within a logarithmic factor of the $O(n\log n)$ bound for the unit-disk case~\cite{ref:TkachenkoDo25}.

Both algorithms rely on new structural lemmas that exploit the ordering of disks and reduce the problem to efficiently testing certain configurations. In the center-convex case, the ordering is immediate: it is induced by the cyclic order of the centers along the convex hull of all disk centers. This leads to a key structural lemma (Lemma~\ref{lem:3points}) that directly yields an efficient algorithm. In the disk-convex case, defining such an order is more subtle, since the boundary of a single disk may appear on the convex hull of all disks in multiple disjoint arcs. We resolve this difficulty by introducing a suitable ordering and proving an analogue of the key lemma (Lemma~\ref{lem:key}). The proof is significantly more involved and constitutes one of the main technical contributions of this paper.

To the best of our knowledge, this is the first systematic study of independent sets of size three in disk graphs. Our results show that even this smallest nontrivial case already exhibits rich structural properties, particularly under convexity assumptions, and that these properties can be exploited algorithmically. We believe that the techniques developed in this paper may be useful for studying larger independent sets and related problems in geometric intersection graphs.

\paragraph{Outline.}
The rest of the paper is organized as follows. In Section~\ref{sec:pre}, we introduce notation and basic definitions. Sections~\ref{sec:3IScntr-cnvx} and~\ref{sec:disconvex} present our algorithms for the center-convex and disk-convex cases, respectively. The proof of the key lemma (Lemma~\ref{lem:key}) for the disk-convex case, which is quite long and technical, is deferred to Section~\ref{sec:lemkey}. Finally, in Section~\ref{sec:general}, we present our algorithm for the general case.

\section{Preliminaries}
\label{sec:pre}

Let $\cd$ be a set of $n$ disks in the plane. For each disk $D_i \in \cd$, let $p_i$ denote its center and $r_i$ its radius. Let $P = \{p_i \mid D_i \in \cd\}$ be the set of centers.

For any two disks $D_i, D_j \in \cd$, we say that they \emph{intersect} if $|p_i p_j| \le r_i + r_j$, and are \emph{disjoint} otherwise.

For any two points $p$ and $q$ in the plane, let $\overline{pq}$ denote the line segment connecting them, $|pq|$ their Euclidean distance, and $\ell(\overline{pq})$ the supporting line of $\overline{pq}$.

Let $\ch(\cd)$ denote the convex hull of all disks in $\cd$. The boundary of $\ch(\cd)$ consists of an alternating sequence of $O(n)$ circular arcs and line segments~\cite{ref:RappaportA92}, where each line segment is an outer common tangent of two consecutive disks along $\partial \ch(\cd)$.

For any region $R$ in the plane, let $\partial R$ denote its boundary. In particular, $\partial \ch(\cd)$ is the boundary of $\ch(\cd)$, and $\partial D_i$ is the boundary circle of a disk $D_i$.

If two disks $D_i, D_j \in \cd$ satisfy $D_i \subseteq D_j$, then $D_j$ can be ignored when computing an independent set of $\cd$. Indeed, any independent set containing $D_j$ can be transformed into one of the same size by replacing $D_j$ with $D_i$. Hence, we assume throughout that no disk in $\cd$ contains another.

We say that $\cd$ is in \emph{disk-convex position} if every disk contributes to $\partial \ch(\cd)$; see Figure~\ref{fig:convexpos}. Under the above assumption, each disk $D_i \in \cd$ contains at least one point on $\partial \ch(\cd)$ that is not contained in any other disk.

We say that $\cd$ is in \emph{center-convex position} if every point in $P$ is a vertex of the convex hull of $P$; see Figure~\ref{fig:centerconvex}.


The {\em disk graph} induced by $\cd$ is the graph whose vertex set is $\cd$ and two disks have an edge if they intersect. An {\em independent set} of the graph is a subset of vertices such that no two vertices in the subset have an edge. Alternatively, we say that a subset of disks of $\cd$ is an {\em independent set} if every two disks in the subset are disjoint. 

The following observation, which is self-evident, will be frequently used. 

\begin{observation}\label{obser:10}
Two disks $D_i$ and $D_j$ are disjoint if the segment $\overline{p_i p_j}$ contains a point not in $D_i \cup D_j$.
\end{observation}





\section{The center-convex position case}
\label{sec:3IScntr-cnvx}

We first consider the case where the set $P$ of disk centers is in convex position. Our goal is to compute an independent set of $\cd$ of size $3$. We present an $O(n \log^2 n)$ time algorithm for this case. The algorithm for the disk-convex position case (Section~\ref{sec:disconvex}) follows the same overall framework, but relies on a different proof of a key technical lemma.

Let $\ch(P)$ denote the convex hull of $P$. Since $P$ is in convex position, every point of $P$ is a vertex of $\ch(P)$. Let $\langle p_1, p_2, \ldots, p_n \rangle$ be the cyclic ordering of the points in $P$ along $\partial \ch(P)$ in counterclockwise order.

For any $i, j \in [1,n]$, let $P[i,j]$ denote the contiguous subsequence of $P$ from $p_i$ to $p_j$ in counterclockwise order, inclusive. That is,
\[
P[i,j]=
\begin{cases}
\langle p_i,p_{i+1},\ldots,p_j\rangle, & \text{if } i\le j,\\[2mm]
\langle p_i,p_{i+1},\ldots,p_n,p_1,\ldots,p_j\rangle, & \text{if } i>j.
\end{cases}
\]
If $i = j$, then define $P[i,j] = \langle p_i \rangle$. We also use $P(i, j)$ to denote the subsequence $P[i, j]$ excluding its endpoints $p_i$ and $p_j$.

To exploit this cyclic order, we define, for each point $p_i \in P$, two indices that capture the nearest disjoint disks in each direction.

\begin{definition}
\label{def:ab-cntr}
For a point $p_i \in P$, let $i_1$ (resp., $i_2$) be the index of the first vertex of $\ch(P)$ encountered counterclockwise (resp., clockwise) from $p_i$ whose disk is disjoint from $D_i$. If $D_i$ intersects every disk in $\cd$, then we set $i_1 = i_2 = 0$.
\end{definition}

\subsection{A key lemma}

The following Lemma~\ref{lem:3points} is central to our approach. Given this result, the overall algorithmic framework closely follows that for the \emph{unit-disk} case (where all disks have equal radii) presented in~\cite{ref:TkachenkoDo25}. Our proof generalizes the corresponding argument for unit disks from~\cite{ref:TkachenkoDo25}, which in turn builds on~\cite{ref:KobayashiAn22}.

The algorithm in Section~\ref{sec:disconvex} also follows the same framework. However, the proof of its key lemma (Lemma~\ref{lem:key}) is considerably more involved. In this sense, the proof of Lemma~\ref{lem:3points} can be viewed as a warm-up for the more technical arguments required later.

\begin{lemma}[Key Lemma]
\label{lem:3points}
The set $\cd$ contains an independent set of size $3$ if and only if there exists a point $p_i \in P$ such that the subsequence $P[i_1,i_2]$ contains two centers whose corresponding disks are disjoint.
\end{lemma}

\begin{proof}
We first prove necessity. Let $D_i, D_j, D_k$ be three pairwise disjoint disks whose centers $p_i, p_j, p_k$ appear along $\partial \ch(P)$ in counterclockwise order. Since $D_i$ and $D_j$ are disjoint, it follows from the definition of $i_1$ and $i_2$ that $p_j \in P[i_1,i_2]$. Similarly, $p_k \in P[i_1,i_2]$. As $D_j$ and $D_k$ are also disjoint, the subsequence $P[i_1,i_2]$ contains two centers whose corresponding disks are disjoint. This establishes the necessity.

We now prove sufficiency. Suppose there exists a point $p_i \in P$ such that $P[i_1,i_2]$ contains two centers $p_j$ and $p_k$ whose disks $D_j$ and $D_k$ are disjoint. Note that this implies $i_1 \neq i_2$. Without loss of generality, assume that $p_i, p_j, p_k$ appear along $\partial \ch(P)$ in counterclockwise order. We show that $\cd$ contains an independent set of size $3$ by considering the following cases.

\begin{description}
\item[\bf Case 1.] $D_{i_1}$ and $D_{i_2}$ are disjoint.

In this case, by definition, $D_i$ is disjoint from both $D_{i_1}$ and $D_{i_2}$. Hence, $\{D_i, D_{i_1}, D_{i_2}\}$ is an independent set of size $3$.

\item[\bf Case 2.] $D_{i_1}$ and $D_{i_2}$ intersect, and at least one of $D_j$ and $D_k$ is disjoint from at least one of $D_{i_1}$ and $D_{i_2}$. 

Without loss of generality, assume that $D_j$ is disjoint from $D_{i_2}$. We will prove that $\{D_i, D_j, D_{i_2}\}$ is an independent set.

\item[\bf Case 3.] $D_{i_1}$ and $D_{i_2}$ intersect, and each of $D_j$ and $D_k$ intersects both $D_{i_1}$ and $D_{i_2}$. 

In this case, we will prove that $\{D_i, D_j, D_k\}$ is an independent set. 
\end{description}

We next present the proof details of Cases~2 and~3.

\paragraph{Proving Case 2.}
Without loss of generality, assume that $D_j$ is disjoint from $D_{i_2}$. We show that $\{D_i, D_j, D_{i_2}\}$ is an independent set. Since $D_i$ is disjoint from $D_{i_2}$ by definition, it suffices to prove that $D_i$ and $D_j$ are disjoint.

Without loss of generality, assume that $\overline{p_{i_1}p_{i_2}}$ is horizontal with $p_{i_1}$ to the left of $p_{i_2}$. Since $P$ is in convex position, $p_j$ lies below the line $\ell(\overline{p_{i_1}p_{i_2}})$, and the segment $\overline{p_i p_j}$ intersects $\overline{p_{i_1}p_{i_2}}$ at a point $q$ (see Figure~\ref{fig:cntr_cnvx}). As $D_{i_1}$ and $D_{i_2}$ intersect, we have $\overline{p_{i_1}p_{i_2}} \subseteq D_{i_1} \cup D_{i_2}$, and thus $q \in D_{i_1} \cup D_{i_2}$.

\begin{figure}[t]
\begin{minipage}[t]{0.49\textwidth}
\begin{center}
\includegraphics[height=2.5in]{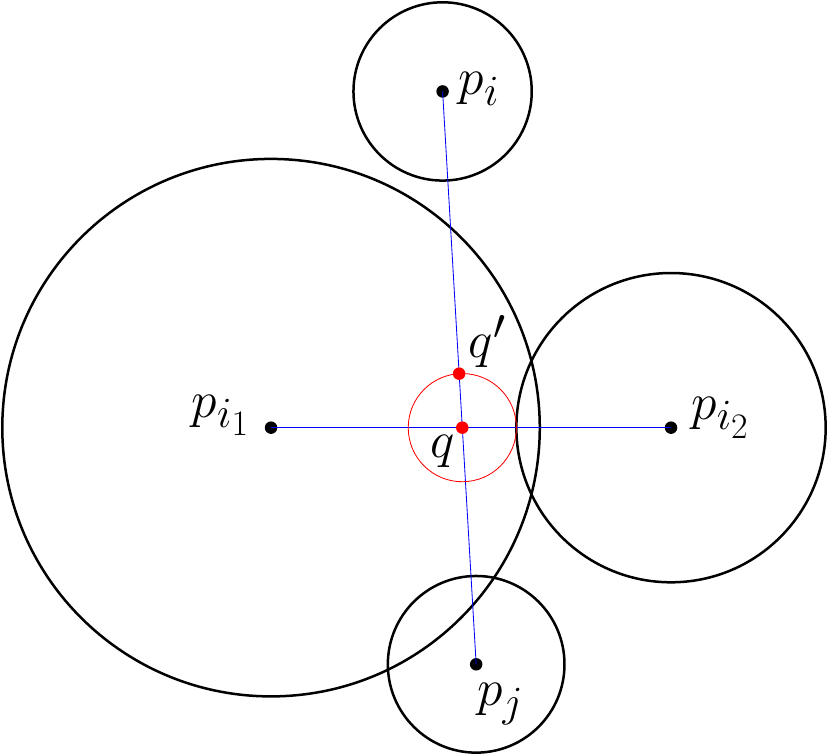}
\caption{\footnotesize Illustrating the proof of Lemma~\ref{lem:3points} for Case 2.}
\label{fig:cntr_cnvx}
\end{center}
\end{minipage}
\hspace{0.02in}
\begin{minipage}[t]{0.49\textwidth}
\begin{center}
\includegraphics[height=1.8in]{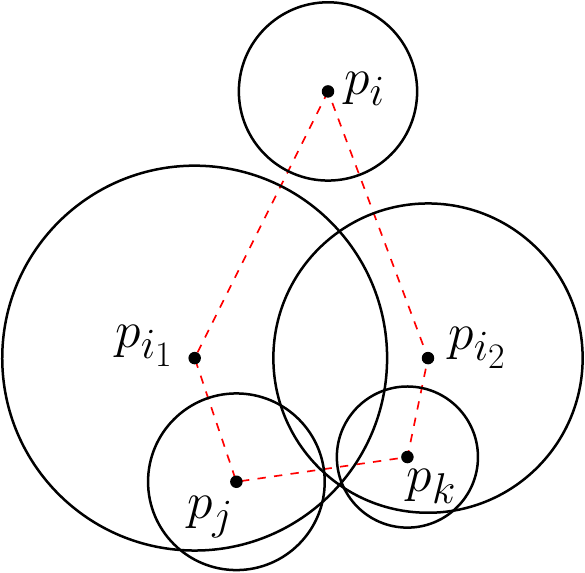}
\caption{\footnotesize Illustrating the proof of Lemma~\ref{lem:3points} for Case 3.}
\label{fig:case3}
\end{center}
\end{minipage}
\vspace{-0.1in}
\end{figure}

We distinguish two subcases depending on whether $D_j$ intersects $D_{i_1}$.

If $D_j$ is disjoint from $D_{i_1}$, then $D_j$ is disjoint from both $D_{i_1}$ and $D_{i_2}$, implying $q \notin D_j$. Similarly, since $D_i$ is disjoint from both $D_{i_1}$ and $D_{i_2}$, we have $q \notin D_i$. As $q \in \overline{p_i p_j}$, it follows from Observation~\ref{obser:10} that $D_i$ and $D_j$ are disjoint.

We now consider the case where $D_j$ intersects $D_{i_1}$.
If $q \in D_{i_2}$, then since $D_{i_2}$ is disjoint from both $D_i$ and $D_j$, we have $q \notin D_i$ and $q \notin D_j$, implying again that $D_i$ and $D_j$ are disjoint. Hence, assume that $q \notin D_{i_2}$ (see Figure~\ref{fig:cntr_cnvx}).

Let $D_q$ be the disk centered at $q$ with radius equal to the distance from $q$ to $D_{i_2}$; thus $D_q$ is tangent to $D_{i_2}$. Since $D_{i_1} \cap D_{i_2} \neq \emptyset$, it follows that $D_q \subseteq D_{i_1}$. As $D_i$ is disjoint from $D_{i_1}$, it is also disjoint from $D_q$.

Let $q'$ be the intersection point of $\partial D_q$ with $\overline{p_i p_j}$ that lies above $\ell(\overline{p_{i_1}p_{i_2}})$. Since $q' \in D_q$ and $D_i$ is disjoint from $D_q$, we have $q' \notin D_i$.

We claim that $q' \notin D_j$. Suppose for contradiction that $q' \in D_j$. By construction, $q'$ is the point of $D_q$ farthest from $p_j$. Hence, $D_q \subseteq D_j$. Since $D_q$ is tangent to $D_{i_2}$, it intersects $D_{i_2}$, which implies that $D_j$ intersects $D_{i_2}$. This is a contradiction to the assumption that $D_j$ is disjoint from $D_{i_2}$. Therefore, $q' \notin D_j$.

Since $q' \notin D_i$, $q' \notin D_j$, and $q' \in \overline{p_i p_j}$, Observation~\ref{obser:10} implies that $D_i$ and $D_j$ are disjoint.

We conclude that $\{D_i, D_j, D_{i_2}\}$ is an independent set in all cases.

\paragraph{Proving Case 3.}
In this case, each of $D_j$ and $D_k$ intersects both $D_{i_1}$ and $D_{i_2}$. Since $D_j$ and $D_k$ are disjoint while $D_{i_1}$ and $D_{i_2}$ intersect, it follows that $p_j, p_k \in P(i_1,i_2)$.

We show that $\{D_i, D_j, D_k\}$ is an independent set. By the convex position of $P$, the points $p_i, p_{i_1}, p_j, p_k, p_{i_2}$ appear in this order along $\partial \ch(P)$ and form a convex pentagon (see Figure~\ref{fig:case3}). In particular, the segments $\overline{p_i p_j}$ and $\overline{p_{i_1} p_k}$ intersect.
By the triangle inequality applied to this crossing, we obtain
\[
|p_i p_j| + |p_{i_1} p_k| \ge |p_j p_k| + |p_i p_{i_1}|.
\]
Since $D_{i_1}$ intersects $D_k$, we have $|p_{i_1} p_k| \le r_{i_1} + r_k$. Substituting this into the above inequality yields
\begin{equation}\label{equ:10}
|p_i p_j| + r_{i_1} + r_k \ge |p_j p_k| + |p_i p_{i_1}|.
\end{equation}

On the other hand, since $D_j$ and $D_k$ are disjoint and $D_i$ and $D_{i_1}$ are disjoint, we have
\begin{equation}\label{equ:20}
|p_j p_k| + |p_i p_{i_1}| > (r_j + r_k) + (r_i + r_{i_1}).
\end{equation}

Combining \eqref{equ:10} and \eqref{equ:20}, we obtain
\[
|p_i p_j| > r_i + r_j,
\]
which implies that $D_i$ and $D_j$ are disjoint.

By symmetry, the same argument shows that $D_i$ and $D_k$ are also disjoint. Therefore, $\{D_i, D_j, D_k\}$ is an independent set.
\end{proof}

For a point $p_i \in P$, if the subsequence $P[i_1,i_2]$ contains two points whose corresponding disks are disjoint, then we call $p_i$ a \emph{feasible point}. By Lemma~\ref{lem:3points}, computing an independent set of size $3$ reduces to identifying a feasible point. The next section presents an algorithm for finding such a point.

\subsection{Algorithm}

Based on the above discussion, our algorithm consists of two stages. In the first stage, we compute the indices $i_1$ and $i_2$ for all points $p_i \in P$. In the second stage, we use these indices to determine whether a feasible point exists.

For the first stage, we use a result from~\cite{ref:TkachenkoCo26} (see Lemma~10 in the arXiv version) to obtain the following lemma.

\begin{lemma}[\cite{ref:TkachenkoCo26}]
\label{lem:firstout}
We can construct a data structure for $P$ in $O(n \log^2 n)$ time such that, for any point $p_i \in P$, the indices $i_1$ and $i_2$ can be computed in $O(\log^2 n)$ time.
\end{lemma}

\begin{proof}
In the preprocessing phase, we construct a complete binary tree $T$ whose leaves, from left to right, store the points $p_1, p_2, \ldots, p_n$. For each node $v \in T$, we build the farthest-disk Voronoi diagram~\cite{ref:RappaportCo89} for the disks centered at the points in $P_v$, where $P_v$ is the subset of points of $P$ stored in the leaves of the subtree rooted at $v$.

Using this structure, the indices $i_1$ and $i_2$ for any point $p_i$ can be computed in $O(\log^2 n)$ time. We refer to Lemma~10 in the arXiv version of~\cite{ref:TkachenkoCo26} for details.
\end{proof}

By Lemma~\ref{lem:firstout}, the indices $i_1$ and $i_2$ for all points $p_i \in P$ can be computed in a total of $O(n \log^2 n)$ time.

In the second stage, our goal is to find a feasible point, if one exists. To this end, we use the following observation, whose proof is similar to Observation~2 in the arXiv version of~\cite{ref:TkachenkoDo25}.

\begin{observation}[\cite{ref:TkachenkoDo25}]
\label{obser:20}
A point $p_i \in P$ is feasible if and only if there exists a point $p_j \in P[i_1,i_2]$ such that $p_{j_1} \in P[i_1,i_2]$ and the triple $(p_i, p_j, p_{j_1})$ appears in counterclockwise order along $\partial \ch(P)$.
\end{observation}

\begin{proof}
Suppose there exists a point $p_j \in P[i_1,i_2]$ such that $p_{j_1} \in P[i_1,i_2]$ and $(p_i, p_j, p_{j_1})$ appears in counterclockwise order along $\partial \ch(P)$. By the definition of $j_1$, the disks $D_j$ and $D_{j_1}$ are disjoint. Since both $p_j$ and $p_{j_1}$ lie in $P[i_1,i_2]$, it follows that $p_i$ is a feasible point.

Conversely, suppose that $p_i$ is feasible. Then $P[i_1,i_2]$ contains two centers whose corresponding disks are disjoint. Let these centers be $p_j$ and $p_k$, chosen such that $(p_i, p_j, p_k)$ appears in counterclockwise order along $\partial \ch(P)$. By the definition of $j_1$, we have $p_{j_1} \in P[j,k]$. Since $p_k \in P[i_1,i_2]$, it follows that $p_{j_1} \in P[i_1,i_2]$. Moreover, as $p_j \in P[i_1,i_2]$, the triple $(p_i, p_j, p_{j_1})$ appears in counterclockwise order along $\partial \ch(P)$.
\end{proof}


For each point $p_i\in P$, note that $i_1\neq i$. Define
\[
i_1'=
\begin{cases}
i_1, & \text{if } i<i_1,\\
i_1+n, & \text{otherwise.}
\end{cases}
\]

Note that if $i_1=i_2$, then the subsequence $P[i_1,i_2]$ contains only one point, and hence $p_i$ cannot be feasible. Therefore, we only need to consider points $p_i$ with $i_1\neq i_2$.
The following lemma, which is based on Observation~\ref{obser:20}, is proved in exactly the same way as in \cite{ref:TkachenkoDo25} for the unit-disk case. 

\begin{lemma}[\cite{ref:TkachenkoDo25}]\label{lem:feasiblepoint}
For each point $p_i\in P$, the following hold.
\begin{enumerate}
    \item If $i_1<i_2$, then $p_i$ is feasible if and only if $\min_{j\in [i_1,i_2]} j_1' \le i_2$.

    \item If $i_1>i_2$, then $p_i$ is feasible if and only if
    $\min_{j\in [i_1,n]} j_1' \le i_2+n$
    or
    $\min_{j\in [1,i_2]} j_1' \le i_2$.

\end{enumerate}
\end{lemma}
\begin{proof}
See Lemma~17 in the arXiv version of \cite{ref:TkachenkoDo25} for the proof. 
\end{proof}

In light of Lemma~\ref{lem:feasiblepoint}, we can identify a feasible point as follows. Define an array $A[1 \cdots n]$ by setting $A[j] = j_1'$ for each $1 \le j \le n$. By Lemma~\ref{lem:feasiblepoint}, for each point $p_i \in P$, we can test whether $p_i$ is feasible using at most two \emph{range-minimum queries} of the form: given an interval $[i',j']$ with $i' \le j'$, return
\[
\min \{ A[i'], A[i'+1], \ldots, A[j'] \}.
\]

Such queries can be answered in $O(1)$ time after $O(n)$ time preprocessing on $A$~\cite{ref:BenderTh00,ref:HarelFa84}. For our purpose, it suffices to use a simpler data structure with $O(\log n)$ query time and $O(n \log n)$ preprocessing time, which can be implemented using an augmented binary search tree. Therefore, once the array $A$ is constructed, a feasible point (if one exists) can be found in $O(n \log n)$ time.

Combining this with the preprocessing of Lemma~\ref{lem:firstout}, we obtain an $O(n \log^2 n)$ time algorithm to decide whether $\cd$ contains an independent set of size $3$. If such a set exists, it can be recovered within the same time bound using the arguments in Lemma~\ref{lem:3points} and Observation~\ref{obser:20}. We thus obtain the following theorem.

\begin{theorem}
\label{theo:centerconvexindset3}
Given a set of $n$ disks in the plane whose centers are in convex position, one can compute in $O(n \log^2 n)$ time a subset of three pairwise disjoint disks, or report that no such subset exists.
\end{theorem}

\section{The disk-convex position case}
\label{sec:disconvex}

We now turn to the disk-convex position case. In this setting, the disks of $\cd$ are in convex position (see Figure~\ref{fig:convexpos}). Our goal is to compute an independent set of size $3$.

We follow the same algorithmic framework as in Section~\ref{sec:3IScntr-cnvx}. The main challenge is to define a suitable cyclic order of the disks. In the center-convex case, this order is induced directly by the cyclic order of the centers along $\partial \ch(P)$. In the disk-convex case, however, such an ordering is less straightforward. A natural approach is to follow the order in which disks appear along $\partial \ch(\cd)$. The difficulty is that a single disk may contribute multiple disjoint arcs to $\partial \ch(\cd)$ and thus may appear multiple times in this traversal. We resolve this issue as follows.

\paragraph{Defining an order for disks.}
We first construct the convex hull $\ch(\cd)$ of $\cd$, which can be done in $O(n \log n)$ time~\cite{ref:RappaportA92}. The boundary $\partial \ch(\cd)$ consists of $O(n)$ circular arcs and line segments~\cite{ref:RappaportA92}. We traverse $\partial \ch(\cd)$ in counterclockwise order and build a cyclic list $L$ of disk centers as follows: whenever we encounter an arc belonging to a disk $D$, we add its center to $L$ if it has not already been added. In this way, each disk center appears exactly once in $L$.
We index the disks according to this cyclic order as $\langle p_1, p_2, \ldots, p_n \rangle$, and use this indexing throughout the section.

For any disk $D \in \cd$, a point $q \in \partial D$ is called a \emph{hull point} if $q \in \partial \ch(\cd)$ and $q$ is not contained in any other disk of $\cd$. Since the disks are in convex position, every disk has at least one hull point. Moreover, by the construction of the ordering, each disk $D_i$ admits a hull point $q_i$ such that the points $q_1, q_2, \ldots, q_n$ appear along $\partial \ch(\cd)$ in counterclockwise order. This property will be crucial in the analysis.

\paragraph{Algorithm.}
Using the above index order, we define $P[i,j]$ and $P(i,j)$ as in Section~\ref{sec:3IScntr-cnvx}. For each point $p_i \in P$, we define the indices $i_1$ and $i_2$ as in Definition~\ref{def:ab-cntr}. We obtain the following key lemma.

\begin{lemma}[Key Lemma]
\label{lem:key}
The set $\cd$ contains an independent set of size $3$ if and only if there exists a point $p_i \in P$ such that $P[i_1,i_2]$ contains two centers whose corresponding disks are disjoint.
\end{lemma}

The proof of Lemma~\ref{lem:key} is significantly more involved than that of Lemma~\ref{lem:3points} in Section~\ref{sec:3IScntr-cnvx}, and constitutes one of the main technical contributions of this paper. Since the proof is quite long and technical, we defer it to Section~\ref{sec:lemkey}.

Given Lemma~\ref{lem:key}, we define feasible points exactly as in Section~\ref{sec:3IScntr-cnvx}. The indices $i_1$ and $i_2$ can be computed using Lemma~\ref{lem:firstout}, as the construction depends only on the disk ordering. Similarly, Observation~\ref{obser:20} and Lemma~\ref{lem:feasiblepoint} carry over with identical proofs, as they rely solely on the ordering and the definition of feasible points. Consequently, we can apply the same range-minimum query approach to identify a feasible point.

We thus obtain the following result.

\begin{theorem}
\label{theo:diskconvex}
Given a set of $n$ disks in the plane that are in convex position, one can compute in $O(n \log^2 n)$ time a subset of three pairwise disjoint disks, or report that no such subset exists.
\end{theorem}

\section{Proving the key lemma --- Lemma~\ref{lem:key}}
\label{sec:lemkey}

We now prove Lemma~\ref{lem:key}. We follow the notation of Section~\ref{sec:disconvex}, in particular the disk ordering defined there. For simplicity, throughout this section we write $\ch$ for $\ch(\cd)$.

We first prove necessity. The argument is identical to that for Lemma~\ref{lem:3points}. Let $D_i, D_j, D_k$ be three pairwise disjoint disks whose centers $p_i, p_j, p_k$ appear along $\partial \ch$ in counterclockwise order. Since $D_i$ and $D_j$ are disjoint, it follows from the definition of $i_1$ and $i_2$ that $p_j \in P[i_1,i_2]$. Similarly, $p_k \in P[i_1,i_2]$. As $D_j$ and $D_k$ are also disjoint, the subsequence $P[i_1,i_2]$ contains two centers whose corresponding disks are disjoint.

We now prove sufficiency. Suppose that there exists a point $p_i \in P$ such that $P[i_1,i_2]$ contains two centers $p_j$ and $p_k$ whose disks $D_j$ and $D_k$ are disjoint. Note that this implies $i_1 \neq i_2$. Without loss of generality, assume that $p_i, p_j, p_k$ appear along $\partial \ch$ in counterclockwise order. Our goal is to show that $\cd$ contains an independent set of size $3$.

As in the proof of Lemma~\ref{lem:3points}, we distinguish three cases.

\begin{description}
\item[\bf Case 1.] $D_{i_1}$ and $D_{i_2}$ are disjoint.

In this case, by definition, $D_i$ is disjoint from both $D_{i_1}$ and $D_{i_2}$. Hence, $\{D_i, D_{i_1}, D_{i_2}\}$ is an independent set of size $3$.

\item[\bf Case 2.] $D_{i_1}$ and $D_{i_2}$ intersect, and at least one of $D_j$ and $D_k$ is disjoint from at least one of $D_{i_1}$ and $D_{i_2}$.

Without loss of generality, assume that $D_j$ is disjoint from $D_{i_2}$. We will prove that $\{D_i, D_j, D_{i_2}\}$ is an independent set.

\item[\bf Case 3.] $D_{i_1}$ and $D_{i_2}$ intersect, and each of $D_j$ and $D_k$ intersects both $D_{i_1}$ and $D_{i_2}$.

In this case, we will prove that $\{D_i, D_j, D_k\}$ is an independent set. 
\end{description}

It remains to prove Cases~2 and~3. In the following, we assume that $D_{i_1}$ and $D_{i_2}$ intersect. Hence, $\overline{p_{i_1}p_{i_2}} \subseteq D_{i_1} \cup D_{i_2}$.
Without loss of generality, assume that $\overline{p_{i_1}p_{i_2}}$ is horizontal, with $p_{i_1}$ to the left of $p_{i_2}$. For simplicity, we also assume that $D_{i_1}$ and $D_{i_2}$ are not tangent; that is, $\partial D_{i_1} \cap \partial D_{i_2}$ consists of two points. This assumption is only to make the discussion less tedious and does not affect the correctness of the argument. 

Before proceeding, we introduce some additional notation.

\subsection{Notation}
\label{sec:notation}

For any disk $D$ and any two points $q, q' \in \partial D$, let $\partial_{[q,q']} D$ denote the portion of $\partial D$ from $q$ to $q'$ in counterclockwise order.

Since $\overline{p_{i_1}p_{i_2}} \subseteq D_{i_1} \cup D_{i_2} \subseteq \ch$ and $q_{i_1}, q_{i_2} \in \partial \ch$ (recall their definitions in Section~\ref{sec:disconvex}), the concatenation of the three segments
$\overline{q_{i_1}p_{i_1}} \cup \overline{p_{i_1}p_{i_2}} \cup \overline{p_{i_2}q_{i_2}}$
partitions $\ch$ into two regions: one locally above $\overline{p_{i_1}p_{i_2}}$ and the other locally below it (recall that $\overline{p_{i_1}p_{i_2}}$ is horizontal). Let $\ch_1$ denote the upper region and let $\ch_2 = \ch \setminus \ch_1$ denote the lower region. We use $\gamma$ to denote the concatenation of the above three segments.

Since $p_i \in P(i_2,i_1)$ and $D_i$ is disjoint from both $D_{i_1}$ and $D_{i_2}$, it follows that $D_i \subseteq \ch_1$. In particular, $p_i \in \ch_1$.

\begin{figure}[t]
\begin{minipage}[h]{\textwidth}
\begin{center}
\includegraphics[height=3.5in]{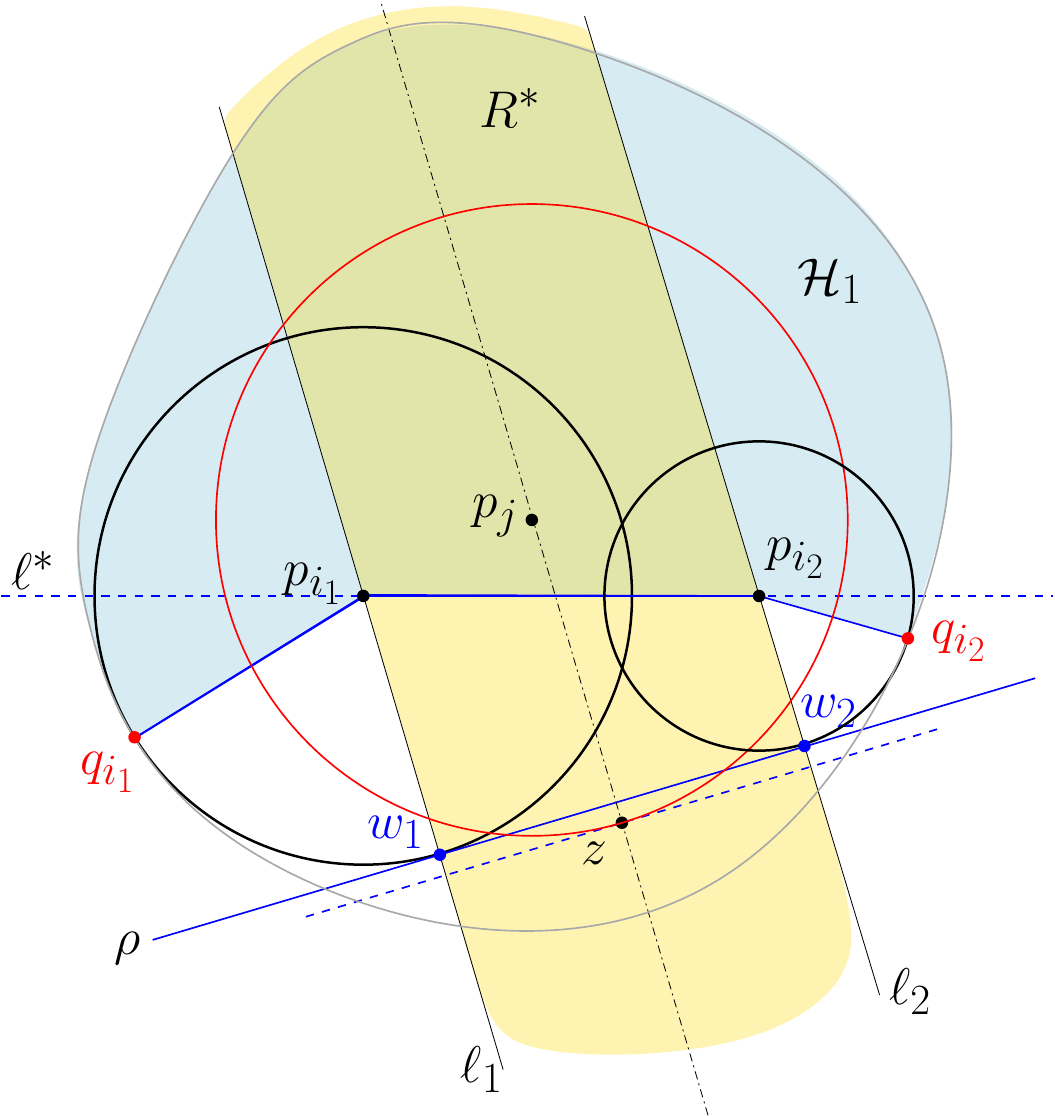}
\caption{\footnotesize Illustration of the notation.}
\label{fig:disk_cnvx}
\end{center}
\end{minipage}
\end{figure}

Recall that $\overline{p_{i_1}p_{i_2}}$ is horizontal.  Let $\ell^*$ refer to $\ell(\overline{p_{i_1}p_{i_2}})$, the supporting line of $\overline{p_{i_1}p_{i_2}}$. Let $\rho$ be the lower common tangent of $D_{i_1}$ and $D_{i_2}$. Let $w_1$ (resp., $w_2$) be the tangency point of $\rho$ with $D_{i_1}$ (resp., $D_{i_2}$); see Figure~\ref{fig:disk_cnvx}. Let $\ell_1$ be the line through $p_{i_1}$ and $w_1$, and note that $\ell_1$ is perpendicular to $\rho$. Similarly, let $\ell_2$ be the line through $p_{i_2}$ and $w_2$. Let $R^*$ be the region bounded by $\ell_1$ and $\ell_2$.

Let the line through $p_j$ parallel to $\ell_1$ intersect $\partial D_j$ at two points, and let $z$ denote the lower intersection point (see Figure~\ref{fig:disk_cnvx}; note that since $\overline{p_{i_1}p_{i_2}}$ is horizontal, $\ell_1$ cannot be horizontal, and thus $z$ is well defined). By definition, $\overline{p_jz}$ is parallel to $\ell_1$ and $\ell_2$. 

Note that since $\overline{p_{i_1}p_{i_2}}$ is horizontal, the common tangent $\rho$ is not vertical. Therefore, it is well-defined to refer to points as lying ``above'' or ``below'' $\rho$ in the following discussion.

The following lemma will be used in the proofs of both Cases~2 and~3. We first apply it to prove Case~2 in Section~\ref{sec:case2}, then prove the lemma in Section~\ref{sec:lem:intersection}, and finally use it again in the proof of Case~3 in Section~\ref{sec:case3}.

\begin{lemma}
\label{lem:intersection}
Assume that $D_{i_1}$ and $D_{i_2}$ intersect. If $p_j \in \ch_1$, then the following hold:
\begin{enumerate}
\item $p_j \in R^*$;
\item $D_j$ intersects $\overline{w_1 w_2}$;
\item $D_{i_1} \cap D_{i_2} \subseteq D_j$.
\end{enumerate}
\end{lemma}

\subsection{Proving Case 2}
\label{sec:case2}

We now prove Case~2. Without loss of generality, assume that $D_j$ is disjoint from $D_{i_2}$. We show that $\{D_i, D_j, D_{i_2}\}$ is an independent set. Since $D_i$ is disjoint from $D_{i_2}$ by definition, it suffices to prove that $D_i$ and $D_j$ are disjoint.

First observe that $p_j \in \ch_2$. Indeed, if $p_j \in \ch_1$, then Lemma~\ref{lem:intersection} would imply that $D_j$ intersects $D_{i_2}$, contradicting the assumption that $D_j$ is disjoint from $D_{i_2}$. Since $p_i \in \ch_1$, the segment $\overline{p_i p_j}$ must intersect $\gamma = \overline{q_{i_1}p_{i_1}} \cup \overline{p_{i_1}p_{i_2}} \cup \overline{p_{i_2}q_{i_2}}$. Therefore, $\overline{p_i p_j}$ intersects at least one of the three segments of $\gamma$.

We consider the following cases.

\begin{enumerate}

\item \textbf{$\boldsymbol{\overline{p_i p_j}}$ intersects $\boldsymbol{\overline{p_{i_2}q_{i_2}}}$.}

Since $\overline{p_{i_2}q_{i_2}} \subseteq D_{i_2}$ and $D_{i_2}$ is disjoint from both $D_i$ and $D_j$, the intersection point lies outside $D_i \cup D_j$. By Observation~\ref{obser:10}, it follows that $D_i$ and $D_j$ are disjoint.

\item \textbf{$\boldsymbol{\overline{p_i p_j}}$ intersects $\boldsymbol{\overline{p_{i_1}q_{i_1}}}$.}

Let $q$ be the point in $\partial D_{i_1} \cap \overline{p_i p_j}$ closest to $p_i$; see Figure~\ref{fig:case2-10}. Then $|p_{i_1} q| = r_{i_1}$ and the segment $\overline{p_j q}$ intersects $\overline{p_{i_1}q_{i_1}}$. Since $D_i$ is disjoint from $D_{i_1}$, we have $q \notin D_i$.

We claim that $q \notin D_j$. Suppose, for contradiction, that $q \in D_j$. Since $\overline{p_j q}$ intersects $\overline{p_{i_1}q_{i_1}}$, the triangle inequality gives
\[
|p_j q| + |p_{i_1} q_{i_1}| \ge |p_j q_{i_1}| + |p_{i_1} q|.
\]
Using $|p_{i_1} q_{i_1}| = |p_{i_1} q| = r_{i_1}$, we obtain $|p_j q| \ge |p_j q_{i_1}|$. Since $q \in D_j$, this implies $q_{i_1} \in D_j$, which contradicts the fact that $q_{i_1}$ lies on $\partial \ch$ and is not contained in any disk other than $D_{i_1}$. Hence $q \notin D_j$.

Since $q \notin D_i \cup D_j$ and $q \in \overline{p_i p_j}$, Observation~\ref{obser:10} implies that $D_i$ and $D_j$ are disjoint.

\begin{figure}[t]
\begin{minipage}[h]{0.49\textwidth}
\begin{center}
\includegraphics[height=3in]{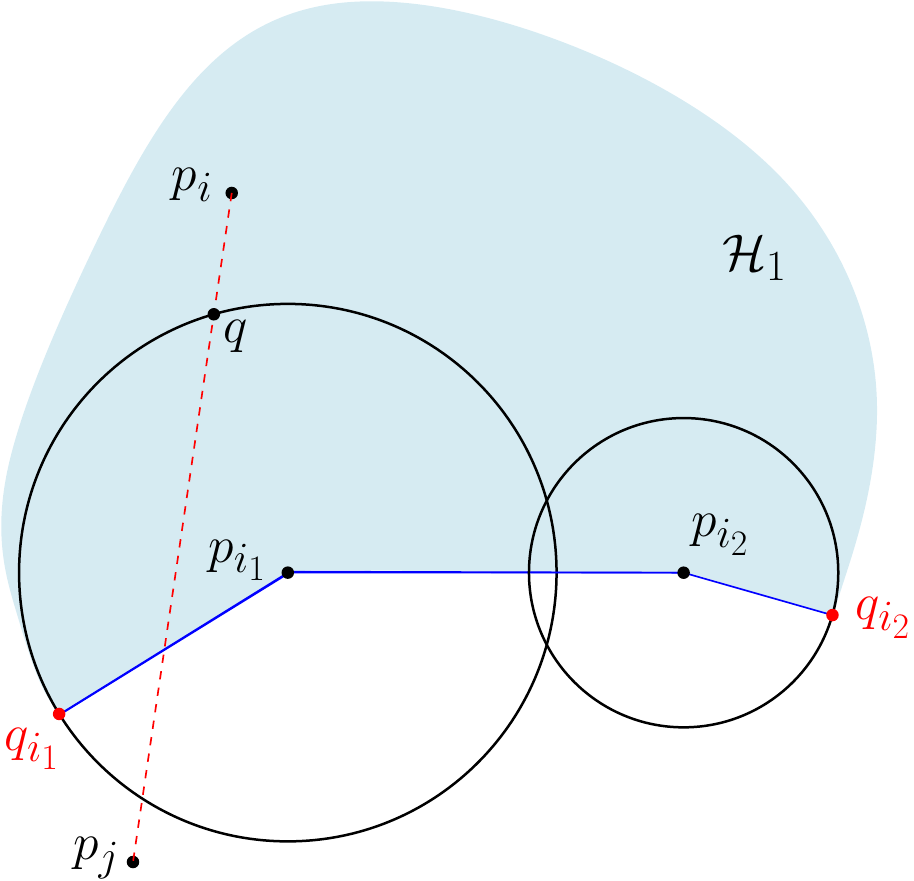}
\caption{\footnotesize Illustrating the proof of Case 2 when $\overline{p_ip_j}$ intersects $\overline{p_{i_1}q_{i_1}}$.}
\label{fig:case2-10}
\end{center}
\end{minipage}
\begin{minipage}[h]{0.49\textwidth}
\begin{center}
\includegraphics[height=3in]{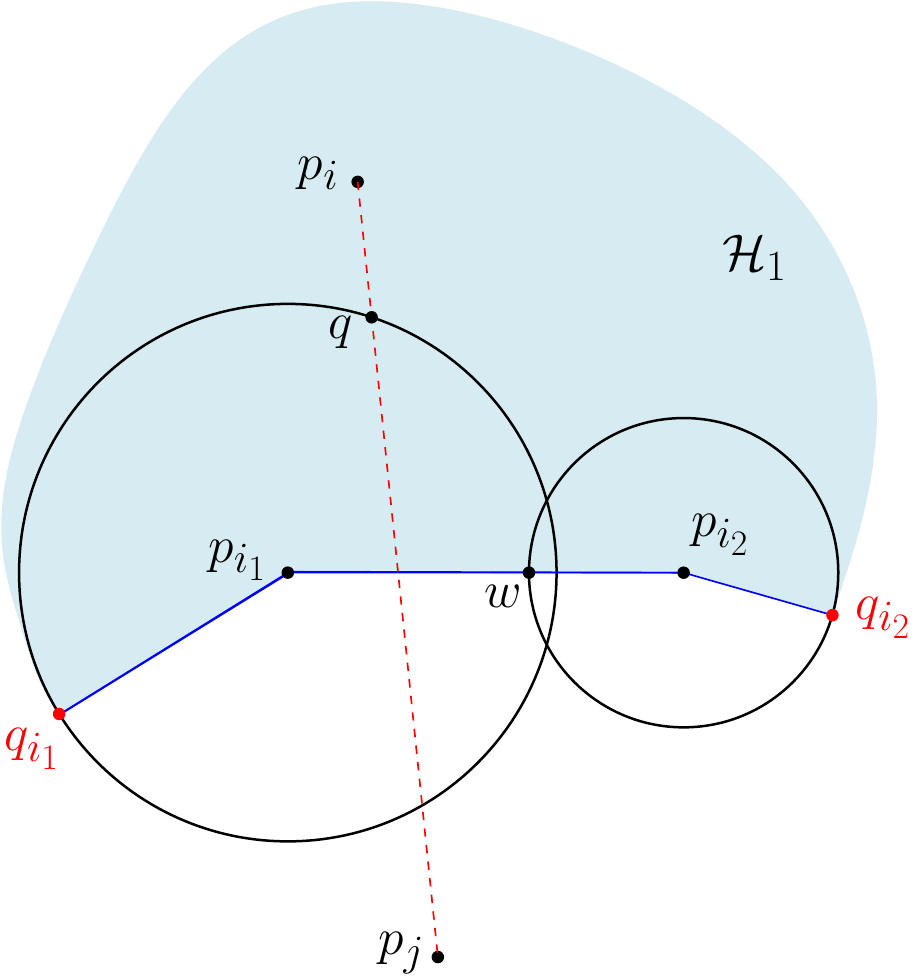}
\caption{\footnotesize Illustrating the proof of Case 2 when $\overline{p_ip_j}$ intersects $\overline{p_{i_1}p_{i_2}}$.}
\label{fig:case2-20}
\end{center}
\end{minipage}
\vspace{-0.1in}
\end{figure}

\item \textbf{$\boldsymbol{\overline{p_i p_j}}$ intersects $\boldsymbol{\overline{p_{i_1}p_{i_2}}}$.}

We distinguish two subcases depending on whether $p_{i_1} \in D_{i_2}$.

If $p_{i_1} \in D_{i_2}$, then  $\ol{p_{i_1}p_{i_2}}\subseteq D_{i_2}$. As $D_{i_2}$ is disjoint from both $D_i$ and $D_j$, the intersection  $\ol{p_ip_j}\cap \ol{p_{i_1}p_{i_2}}$ is not in $D_i\cup D_j$. It follows that $D_i$ and $D_j$ are disjoint by Observation~\ref{obser:10}.

    
If $p_{i_1} \notin D_{i_2}$, then $\overline{p_{i_1}p_{i_2}}$ intersects $\partial D_{i_2}$ at a point $w$ (see Figure~\ref{fig:case2-20}). If $\overline{p_i p_j}$ intersects $\overline{w p_{i_2}}$,  then again $D_i$ and $D_j$ are disjoint following a similar analysis as above. 

Otherwise, $\overline{p_i p_j}$ intersects $\overline{p_{i_1} w}$. Let $q$ be the point in $\partial D_{i_1} \cap \overline{p_i p_j}$ closest to $p_i$ (see Figure~\ref{fig:case2-20}). Then $\overline{q p_j}$ intersects $\overline{p_{i_1} w}$. By the triangle inequality,
\[
|q p_j| + |p_{i_1} w| \ge |p_j w| + |p_{i_1} q|.
\]
Since $\ol{p_{i_1}p_{i_2}}\subseteq D_{i_1}\cup D_{i_2}$, by the definition of $w$, we have $w\in D_{i_1}$. Hence, $|p_{i_1}w|\le r_{i_1}$. Since $q\in \partial D_{i_1}$, $|p_{i_1}q|=r_{i_1}$. Thus, we obtain $|qp_j|\ge |p_jw|$. 

We claim that $q \notin D_j$. Otherwise, since $|qp_j|\ge |p_jw|$, we have $w \in D_j$, which contradicts the assumption that $D_j$ and $D_{i_2}$ are disjoint (since $w \in D_{i_2}$). Thus $q \notin D_j$.

Since $q \in D_{i_1}$ and $D_{i_1}$ is disjoint from $D_i$, we have $q \notin D_i$. Therefore, $q \notin D_i \cup D_j$, and Observation~\ref{obser:10} implies that $D_i$ and $D_j$ are disjoint since $q\in \ol{p_ip_j}$.

\end{enumerate}

In all cases, we have $D_i \cap D_j = \emptyset$. Therefore, $\{D_i, D_j, D_{i_2}\}$ is an independent set. This completes the proof of Case~2.

\begin{figure}[t]
\begin{minipage}[t]{0.49\textwidth}
\begin{center}
\includegraphics[height=3in]{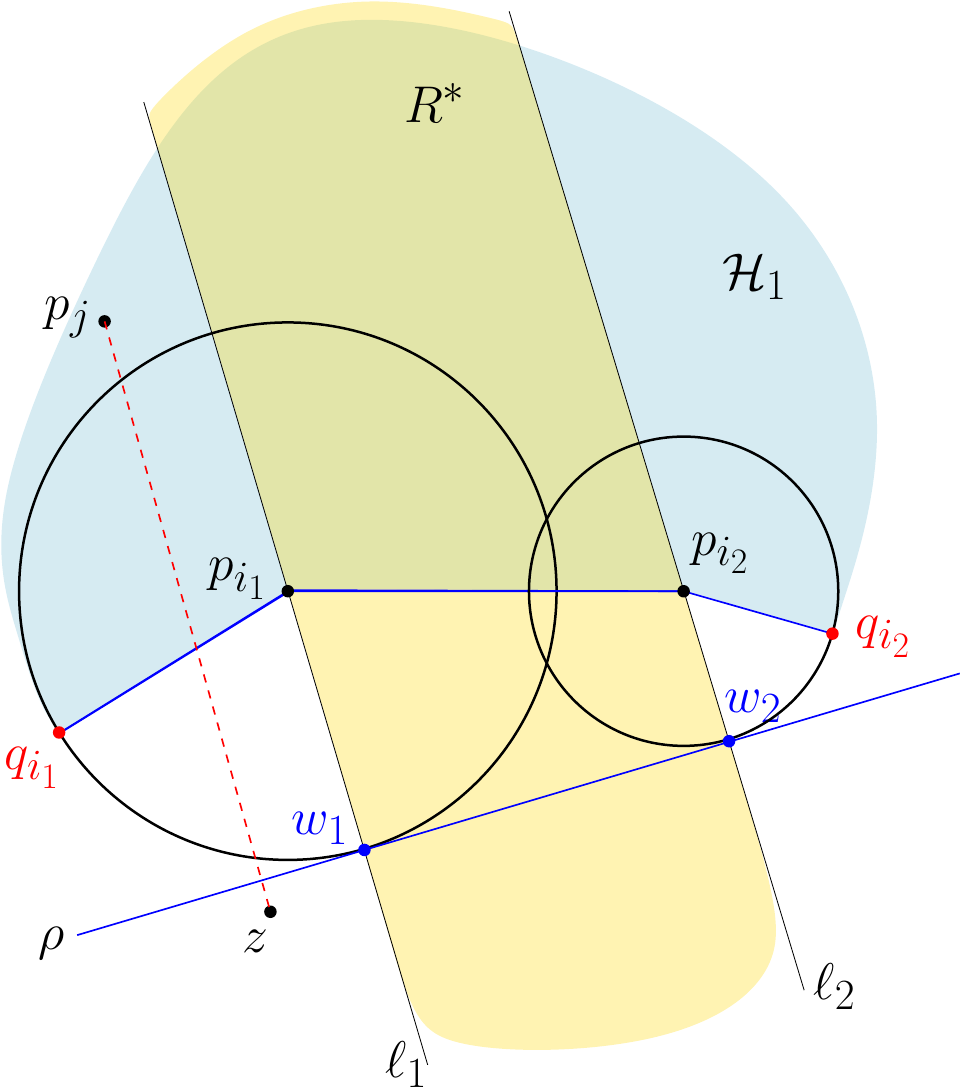}
\caption{\footnotesize Illustration of the proof of Lemma~\ref{lem:intersection}.}
\label{fig:case2-30}
\end{center}
\end{minipage}
\begin{minipage}[t]{0.49\textwidth}
\begin{center}
\includegraphics[height=3in]{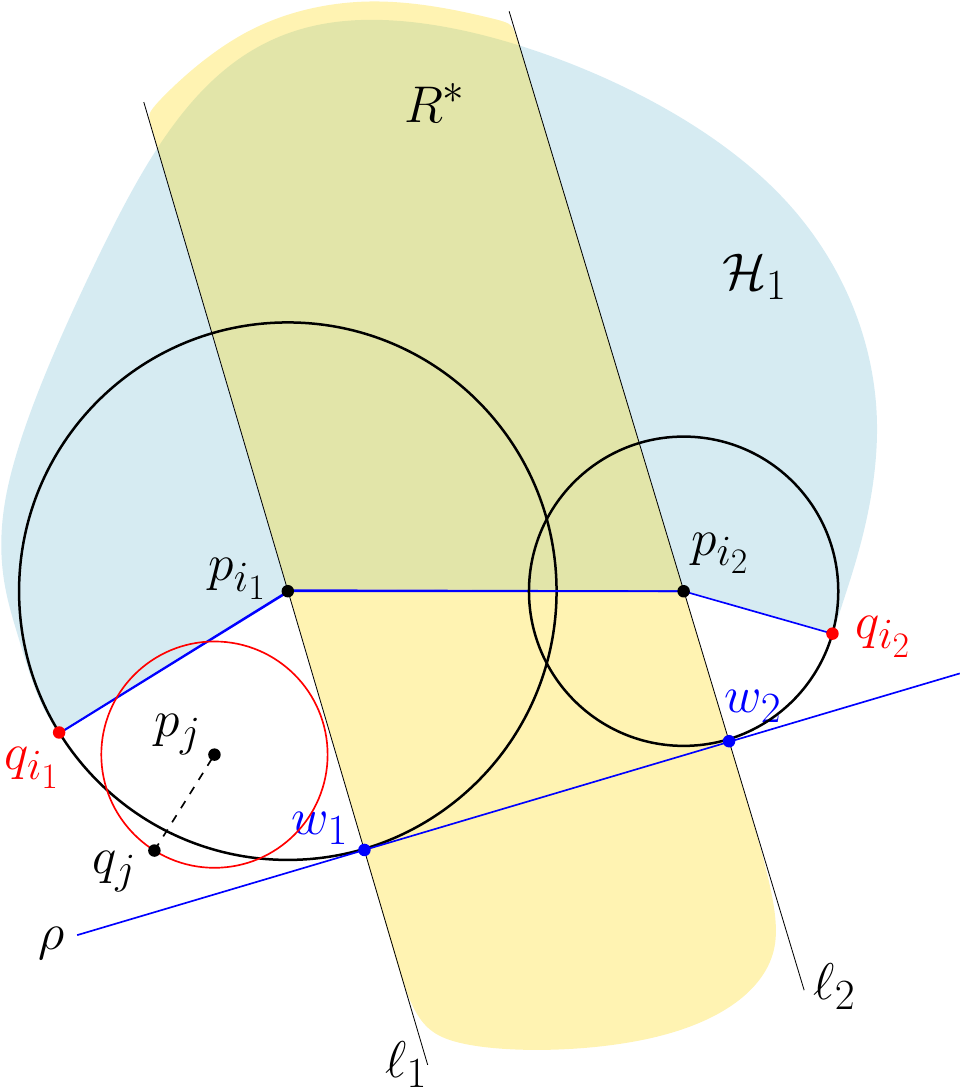}
\caption{\footnotesize Illustration of the proof of Lemma~\ref{lem:intersection}.}
\label{fig:case2-40}
\end{center}
\end{minipage}
\vspace{-0.1in}
\end{figure}

\subsection{Proving Lemma~\ref{lem:intersection}}
\label{sec:lem:intersection}

We now prove Lemma~\ref{lem:intersection}. First, observe that if $j = i_1$ or $j = i_2$, then all three statements of the lemma hold trivially. Hence, in the following, we assume that $j \notin \{i_1,i_2\}$. We prove the three statements in order.

\paragraph{Proving $\boldsymbol{p_j \in R^*}$.}

Assume to the contrary that $p_j \notin R^*$. Then $p_j$ lies either to the left of $\ell_1$ or to the right of $\ell_2$. We assume the former case, as the latter can be treated analogously.

We first argue that $D_j$ does not intersect $\overline{w_1 w_2}$. Indeed, suppose for contradiction that $D_j$ intersects $\overline{w_1 w_2}$. Then the point $z$ must lie below $\rho$ (recall the definition of $z$ in Section~\ref{sec:notation}; see Figure~\ref{fig:case2-30}). By definition, $\overline{p_j z}$ is parallel to $\ell_1$. Since $p_j$ is to the left of $\ell_1$, the point $z$ is also to the left of $\ell_1$.

Furthermore, since $p_j$ is to the left of $\ell_1$ and $p_j \in \ch_1$, the point $q_{i_1}$ must lie to the left of $p_{i_1}$. In addition, since $\overline{p_{i_1} q_{i_1}} \subseteq D_{i_1}$, $D_{i_1}$ lies above $\rho$, and $z$ lies below $\rho$, the segment $\overline{p_j z}$ must intersect $\overline{p_{i_1} q_{i_1}}$ (see Figure~\ref{fig:case2-30}). Consequently, by the triangle inequality, we have
\[
|p_j z| + |p_{i_1} q_{i_1}| \ge |p_{i_1} z| + |p_j q_{i_1}|.
\]
Since $z$ lies below $\rho$, we have $|p_{i_1} z| \ge r_{i_1}$. Because $|p_{i_1} q_{i_1}| = r_{i_1}$, it follows that $|p_j z| \ge |p_j q_{i_1}|$. Since $z \in D_j$, we obtain $q_{i_1} \in D_j$, which contradicts the fact that $q_{i_1}$ does not belong to any disk of $\cd$ other than $D_{i_1}$. Therefore, $D_j$ cannot intersect $\overline{w_1 w_2}$.

Since $p_j \in P[i_1,i_2]$ and $j \notin \{i_1,i_2\}$, we have $p_j \in P(i_1,i_2)$. By the definitions of $q_{i_1}$, $q_{i_2}$, and $q_j$, and since $p_j \in \ch_1$, the segment $\overline{p_j q_j}$ must intersect one of the following three portions: $\partial_{[q_{i_1}, w_1]} D_{i_1}$, $\overline{w_1 w_2}$, or $\partial_{[w_2, q_{i_2}]} D_{i_2}$. Because $D_j$ does not intersect $\overline{w_1 w_2}$ and $\overline{p_j q_j} \subseteq D_j$, it follows that $\overline{p_j q_j}$ does not intersect $\overline{w_1 w_2}$. Hence, $\overline{p_j q_j}$ intersects either $\partial_{[q_{i_1}, w_1]} D_{i_1}$ or $\partial_{[w_2, q_{i_2}]} D_{i_2}$. We assume the former case, as the latter can be treated analogously.

Since $D_j$ does not intersect $\overline{w_1 w_2}$, we have $w_1 \notin D_j$. By definition, $q_{i_1} \notin D_j$ and $q_j \notin D_{i_1}$. Under these conditions, for $\overline{p_j q_j}$ to intersect $\partial_{[q_{i_1}, w_1]} D_{i_1}$, the point $p_j$ must lie in the interior of the circular sector of $D_{i_1}$ bounded by $\overline{p_{i_1} q_{i_1}}$, $\overline{p_{i_1} w_1}$, and $\partial_{[q_{i_1}, w_1]} D_{i_1}$; see Figure~\ref{fig:case2-40}. Note that every interior point of this region lies outside $\ch_1$. Hence, we obtain $p_j \notin \ch_1$, which contradicts the assumption that $p_j \in \ch_1$.

This proves that $p_j \in R^*$.

\paragraph{Proving that $\boldsymbol{D_j}$ intersects $\boldsymbol{\overline{w_1w_2}}$.}
Following the above argument, since $p_j\in \ch_1$,  $\overline{p_jq_j}$ must intersect $\partial_{[q_{i_1},w_1]}D_{i_1}$, $\overline{w_1w_2}$, or $\partial_{[w_2,q_{i_2}]}D_{i_2}$. The above shows that $\overline{p_jq_j}$ cannot intersect $\partial_{[q_{i_1},w_1]}D_{i_1}$ or $\partial_{[w_2,q_{i_2}]}D_{i_2}$. Hence, $\overline{p_jq_j}$ must intersect $\ol{w_1w_2}$. As $\ol{p_jq_j}\subseteq D_j$, it follows that $D_j$ intersects $\ol{w_1w_2}$. 

\begin{figure}[t]
\begin{minipage}[t]{\textwidth}
\begin{center}
\includegraphics[height=3.5in]{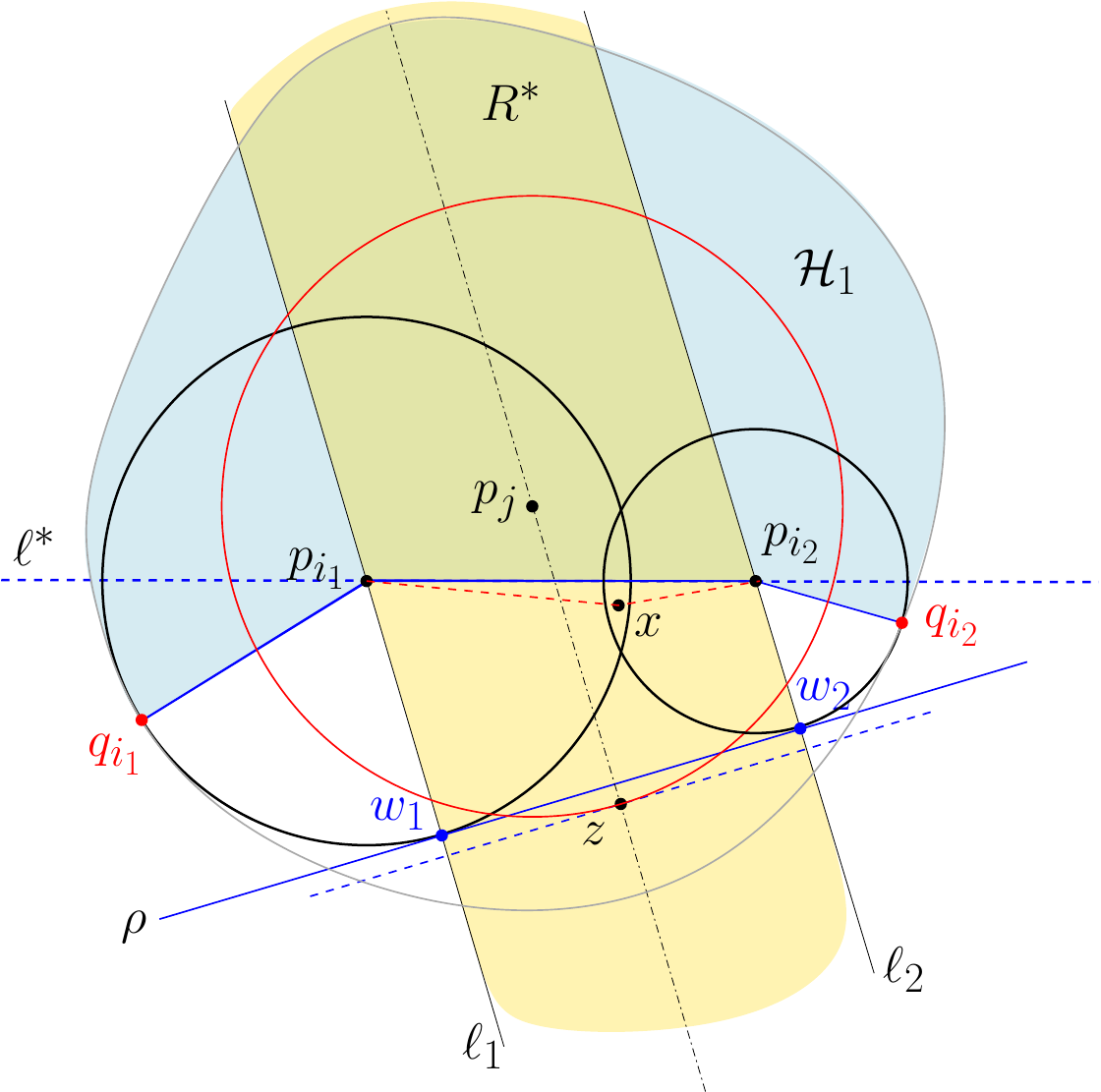}
\caption{\footnotesize Illustration of the proof of Lemma~\ref{lem:intersection}.}
\label{fig:case2-50}
\end{center}
\end{minipage}
\vspace{-0.1in}
\end{figure}

\paragraph{Proving $\boldsymbol{D_{i_1}\cap D_{i_2}\subseteq D_j}$.} 

Since $p_j \in R^*$, $\overline{p_jz}$ is parallel to $\ell_1$, we have $z \in R^*$. Furthermore, because $D_j$ intersects $\overline{w_1 w_2}$ and $p_j \in \ch_1$, the segment $\overline{p_j z}$ must intersect both $\overline{p_{i_1} p_{i_2}}$ and $\overline{w_1 w_2}$. In particular, $z$ cannot lie above $\rho$, and hence $z$ is not in the interior of either $D_{i_1}$ or $D_{i_2}$.

Consider a point $x \in D_{i_1} \cap D_{i_2}$ that lies below the line $\ell^*$ (recall that $\ell^*$ is the supporting line of $\overline{p_{i_1} p_{i_2}}$). Since $\overline{p_j z}$ intersects both $\overline{p_{i_1} p_{i_2}}$ and $\overline{w_1 w_2}$, the segment $\overline{p_j z}$ must intersect either $\overline{p_{i_1} x}$ or $\overline{p_{i_2} x}$ (see Figure~\ref{fig:case2-50}). Without loss of generality, assume that $\overline{p_j z}$ intersects $\overline{p_{i_1} x}$. Then, by the triangle inequality,
\[
|x p_{i_1}| + |p_j z| \ge |p_j x| + |p_{i_1} z|.
\]
Since $x \in D_{i_1}$, we have $|x p_{i_1}| \le r_{i_1}$. Since $z$ is not in the interior of $D_{i_1}$, we have $|p_{i_1} z| \ge r_{i_1}$. Therefore, from the above inequality we obtain $|p_j z| \ge |p_j x|$. As $z \in D_j$, it follows that $x \in D_j$.

Since $x$ is an arbitrary point of $D_{i_1} \cap D_{i_2}$ below $\ell^*$, the above argument shows that all points of $D_{i_1} \cap D_{i_2}$ below $\ell^*$ lie in $D_j$.

Now consider a point $x'$ of $D_{i_1} \cap D_{i_2}$ that lies above $\ell^*$. Let $x$ be the reflection of $x'$ with respect to $\ell^*$. Then $x \in D_{i_1} \cap D_{i_2}$ and lies below $\ell^*$, and hence $x \in D_j$. The line $\ell^*$ is the perpendicular bisector of $x$ and $x'$, and any point above $\ell^*$ is closer to $x'$ than to $x$. Since $p_j \in \ch_1$ and $p_j \in R^*$, we have that $p_j$ lies above $\ell^*$, and thus $|p_j x'| \le |p_j x|$. Since $x \in D_j$, it follows that $x' \in D_j$.

We thus conclude that $D_{i_1} \cap D_{i_2} \subseteq D_j$.

This completes the proof of Lemma~\ref{lem:intersection}.

\subsection{Proving Case 3}
\label{sec:case3}

In this case, each of $D_j$ and $D_k$ intersects both $D_{i_1}$ and $D_{i_2}$. Note that this implies that $p_j, p_k \in P(i_1,i_2)$, since $D_j$ and $D_k$ are disjoint while $D_{i_1}$ and $D_{i_2}$ intersect. We argue that $\{D_i, D_j, D_k\}$ is an independent set. 

We first prove the following lemma. 

\begin{lemma}\label{lem:60}
If $p_k \in \ch_2$ and $p_k$ is strictly above the line $\ell^*$, then $D_k$ contains a point of $D_{i_1} \cap D_{i_2}$. 
\end{lemma}

\begin{proof}
If $p_k \in \ch_2$ and $p_k$ is strictly above $\ell^*$, then either $\overline{p_k p_{i_2}}$ intersects $\overline{p_{i_1} q_{i_1}}$ or $\overline{p_k p_{i_1}}$ intersects $\overline{p_{i_2} q_{i_2}}$. We assume the former case (see Figure~\ref{fig:case3-10}), since the latter can be treated analogously. 

Let $t$ be the intersection point of $\overline{p_k p_{i_2}}$ and $\overline{p_{i_1} q_{i_1}}$. Let $q$ be the intersection point of $\partial D_{i_2}$ with the ray originating from $p_{i_2}$ and passing through $p_k$. We distinguish two cases depending on whether $t \in D_{i_2}$.

\begin{figure}[t]
\begin{minipage}[h]{0.49\textwidth}
\begin{center}
\includegraphics[height=2.0in]{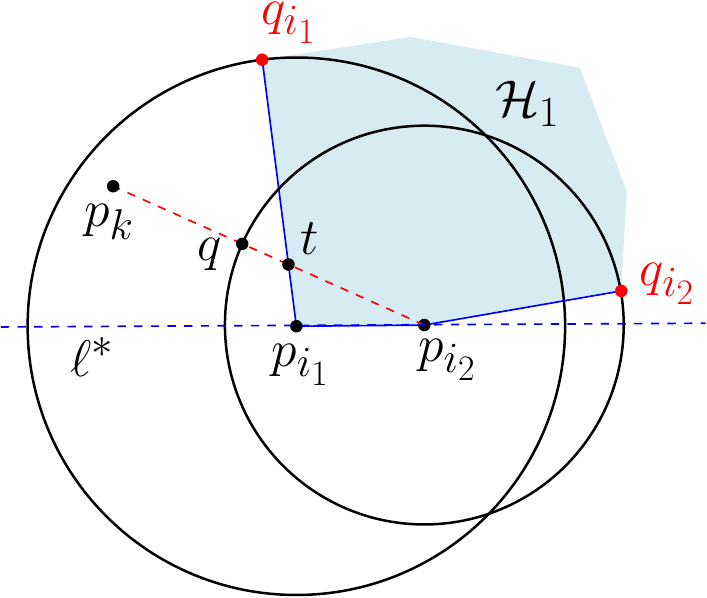}
\caption{\footnotesize Illustrating the proof of Lemma~\ref{lem:60} for the case $t \in D_{i_2}$.}
\label{fig:case3-10}
\end{center}
\end{minipage}
\hspace{0.05in}
\begin{minipage}[h]{0.49\textwidth}
\begin{center}
\includegraphics[height=2.0in]{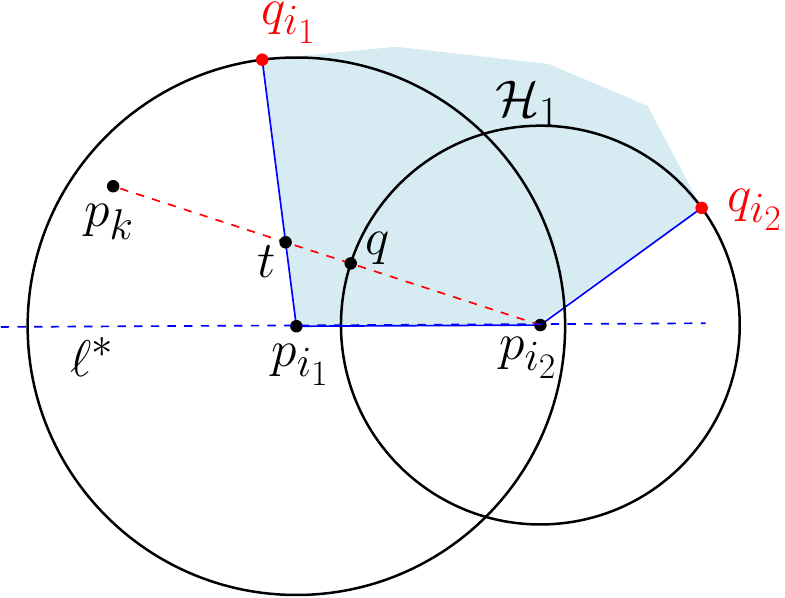}
\caption{\footnotesize Illustrating the proof of Lemma~\ref{lem:60} for the case $t \notin D_{i_2}$.}
\label{fig:case3-20}
\end{center}
\end{minipage}
\vspace{-0.1in}
\end{figure}

\begin{itemize}
    \item If $t \in D_{i_2}$, then $t$ is also the intersection of $\overline{p_{i_1} q_{i_1}}$ and $\overline{p_{i_2} q}$ (see Figure~\ref{fig:case3-10}). By the triangle inequality, we have
    \[
    |p_{i_1} q_{i_1}| + |p_{i_2} q| \ge |p_{i_1} q| + |p_{i_2} q_{i_1}|.
    \]
    By definition, $q_{i_1} \notin D_{i_2}$, and hence $|p_{i_2} q_{i_1}| > r_{i_2}$. Since $q \in \partial D_{i_2}$, we have $|p_{i_2} q| = r_{i_2}$. Therefore, $|p_{i_1} q_{i_1}| > |p_{i_1} q|$. As $|p_{i_1} q_{i_1}| = r_{i_1}$, it follows that $q \in D_{i_1}$. 

    Since $D_k$ intersects $D_{i_2}$ and $q$ is the intersection point of $\partial D_{i_2}$ with the ray emanating from $p_{i_2}$ and passing through $p_k$,
    we have $q \in D_k$. 
    Note also that $q \in D_{i_2}$ by definition. Therefore,
    \[
    q \in D_k \cap D_{i_1} \cap D_{i_2}.
    \]
    Hence, $D_k$ contains a point of $D_{i_1} \cap D_{i_2}$.

    \item If $t \notin D_{i_2}$, then $t$ is also the intersection of $\overline{p_{i_1} q_{i_1}}$ and $\overline{p_k q}$ (see Figure~\ref{fig:case3-20}). By the triangle inequality, we have
    \[
    |p_{i_1} q_{i_1}| + |p_k q| \ge |p_{i_1} q| + |p_k q_{i_1}|.
    \]
    Since $D_k$ intersects $D_{i_2}$ and $q \in \overline{p_k p_{i_2}}$, we have $q \in D_k$, and thus $|p_k q| \le r_k$. By definition, $q_{i_1} \notin D_k$, and hence $|p_k q_{i_1}| > r_k$. Consequently, we obtain $|p_{i_1} q_{i_1}| \ge |p_{i_1} q|$, which implies that $q \in D_{i_1}$ since $q_{i_1} \in \partial D_{i_1}$. Recall that $q \in \partial D_{i_2}$. Therefore,
    \[
    q \in D_k \cap D_{i_1} \cap D_{i_2}.
    \]
\end{itemize}

We conclude that $D_k$ contains a point of $D_{i_1} \cap D_{i_2}$.
\end{proof}

\begin{lemma}
\label{lem:case3-cntr-prty}
Both $p_j$ and $p_k$ lie in $\ch_2$.
\end{lemma}

\begin{proof}
Assume, for contradiction, that the lemma statement is false. Then either both $p_j$ and $p_k$ lie in $\ch_1$, or exactly one of them lies in $\ch_1$.

If both $p_j$ and $p_k$ lie in $\ch_1$, then by Lemma~\ref{lem:intersection}(3), both $D_j$ and $D_k$ contain $D_{i_1} \cap D_{i_2}$, and thus $D_j \cap D_k \neq \emptyset$. This contradicts the assumption that $D_j$ and $D_k$ are disjoint. Hence, it is not possible that both $p_j$ and $p_k$ lie in $\ch_1$.

In the following, we consider the case where exactly one of $p_j$ and $p_k$ lies in $\ch_1$. Without loss of generality, assume that $p_j \in \ch_1$ and $p_k \in \ch_2$.

Since $p_j \in \ch_1$, by Lemma~\ref{lem:intersection}(3), we have $D_{i_1} \cap D_{i_2} \subseteq D_j$. If $p_k$ is strictly above $\ell^*$, then by Lemma~\ref{lem:60}, $D_k$ contains a point of $D_{i_1} \cap D_{i_2}$, and hence $D_k$ cannot be disjoint from $D_j$, a contradiction. Therefore, $p_k$ must lie below $\ell^*$ (including the case $p_k \in \ell^*$).

Since $p_j \in \ch_1$, by Lemma~\ref{lem:intersection}(1), we have $p_j \in R^*$. Also, by Lemma~\ref{lem:intersection}(2), the point $z$ (recall its definition in Section~\ref{sec:notation}) lies below the lower common tangent $\rho$ of $D_{i_1}$ and $D_{i_2}$ (see Figure~\ref{fig:case2-50}). 

Note that $p_k$ lies either to the left or to the right of the line $\ell(\overline{p_j z})$. We assume the former case, as the latter can be treated analogously. Depending on whether the triangle $\triangle p_k p_j p_{i_2}$ contains $z$, we distinguish two cases.

\begin{figure}[t]
\begin{minipage}[h]{0.49\textwidth}
\begin{center}
\includegraphics[height=3.0in]{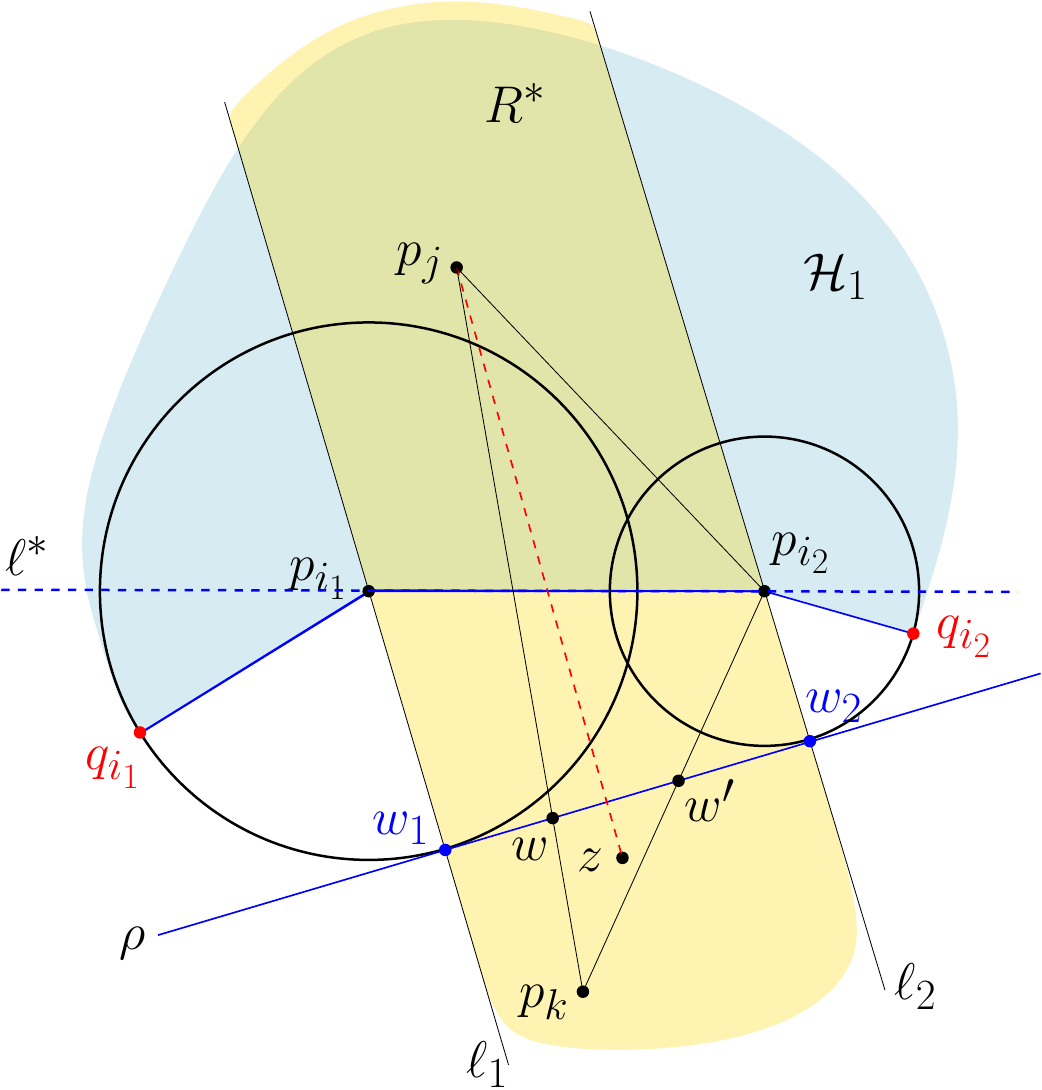}
\caption{\footnotesize Illustrating the proof of Lemma~\ref{lem:case3-cntr-prty} for the case $z \in \triangle p_k p_j p_{i_2}$.}
\label{fig:case3-30}
\end{center}
\end{minipage}
\hspace{0.05in}
\begin{minipage}[h]{0.49\textwidth}
\begin{center}
\includegraphics[height=3.0in]{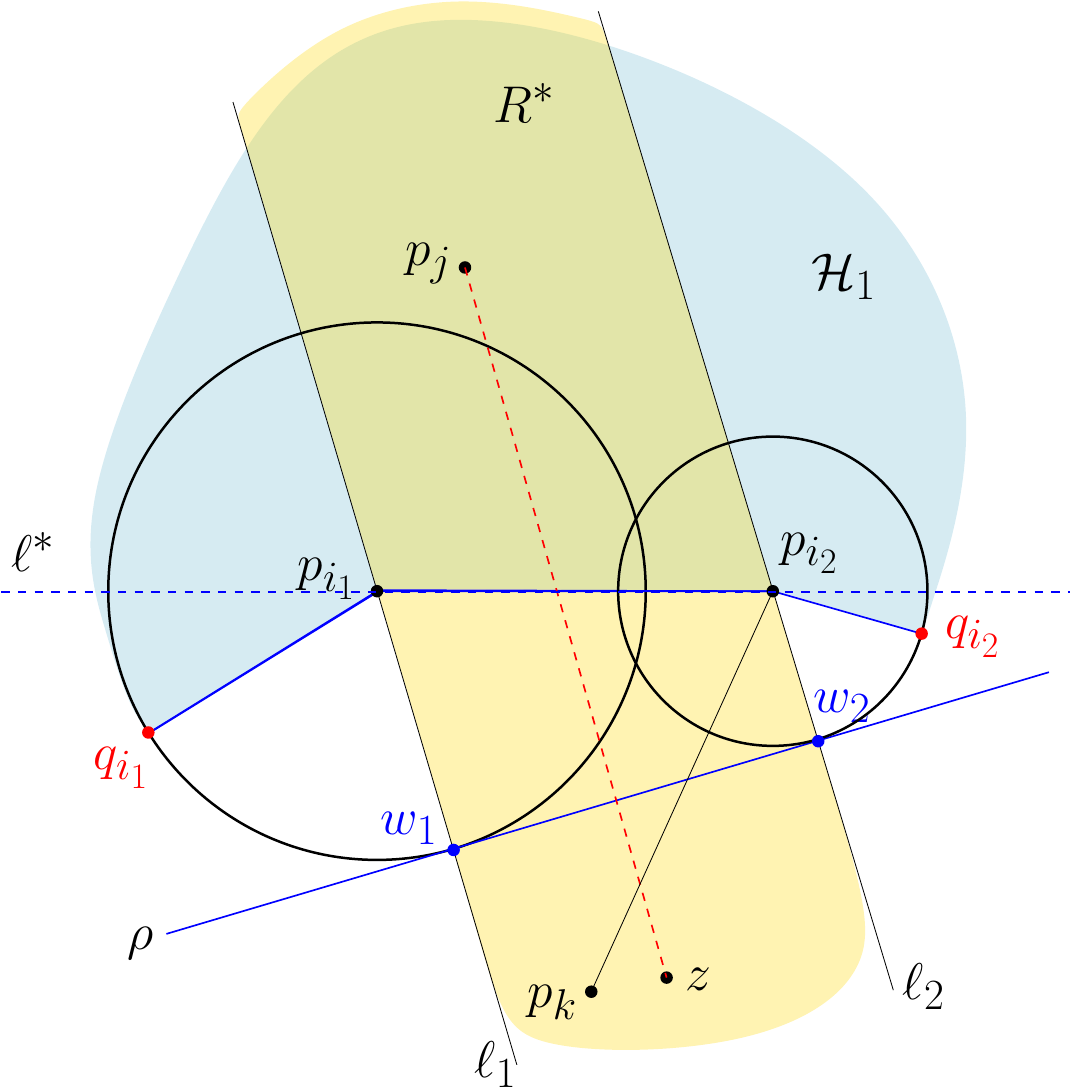}
\caption{\footnotesize Illustrating the proof of Lemma~\ref{lem:case3-cntr-prty} for the case $z \notin \triangle p_k p_j p_{i_2}$.}
\label{fig:case3-40}
\end{center}
\end{minipage}
\vspace{-0.1in}
\end{figure}

\begin{itemize}
\item
If $z \in \triangle p_k p_j p_{i_2}$ (see Figure~\ref{fig:case3-30}), then $z$ lies in the triangle $\triangle p_k w w'$, where $w$ is the intersection of $\rho$ and $\overline{p_j p_k}$, and $w'$ is the intersection of $\rho$ and $\overline{p_{i_2} p_k}$. By definition, $\overline{p_j z}$ is perpendicular to $\overline{w w'}$. Hence, the angle $\angle p_k w w'$ of $\triangle p_k w w'$ is obtuse, and thus $|p_k w'| > |p_k z|$, since $z \in \triangle p_k w w'$.

On the other hand, since $D_k$ and $D_{i_2}$ intersect, $w' \in \overline{p_k p_{i_2}}$, and $w' \notin D_{i_2}$, it follows that $w' \in D_k$. 
Since $|p_k w'| > |p_k z|$, we obtain that $z \in D_k$. As $z$ is also in $D_j$, it follows that $D_k$ intersects $D_j$, a contradiction.

\item
If $z \notin \triangle p_k p_j p_{i_2}$ (see Figure~\ref{fig:case3-40}), then, since $p_k$ and $p_{i_2}$ lie on different sides of $\ell(\overline{p_j z})$, and both $p_k$ and $z$ lie below $\ell^*$ while $p_j$ lies above $\ell^*$, the segment $\overline{p_j z}$ must intersect $\overline{p_k p_{i_2}}$. By the triangle inequality, we have
\[
|p_j z| + |p_k p_{i_2}| \ge |p_j p_k| + |p_{i_2} z|.
\]
By the definition of $z$, we have $|p_j z| = r_j$. Since $D_k$ and $D_{i_2}$ intersect, we have $|p_k p_{i_2}| \le r_k + r_{i_2}$. In addition, since $z$ lies below $\rho$, we have $|p_{i_2} z| \ge r_{i_2}$. Combining these inequalities yields $|p_j p_k| \le r_j + r_k$, implying that $D_j$ and $D_k$ intersect. This is a contradiction, since $D_j$ and $D_k$ are disjoint.
\end{itemize}

We conclude that both cases lead to contradictions. Hence, neither $p_j$ nor $p_k$ can lie in $\ch_1$, and thus both must lie in $\ch_2$. This completes the proof.
\end{proof}

With the above two lemmas, we proceed to prove that $\{D_i, D_j, D_k\}$ is an independent set. 

By Lemma~\ref{lem:case3-cntr-prty}, both $p_j$ and $p_k$ lie in $\ch_2$. We will show that if either $D_j$ or $D_k$ intersects $D_i$, then $D_j$ and $D_k$ cannot be disjoint, which would yield a contradiction since $D_j$ and $D_k$ are disjoint. 

Assume without loss of generality that $D_j$ intersects $D_i$. Our goal is to prove the following proposition. 

\begin{prop}\label{prop:10}
$D_j$ intersects $D_k$.     
\end{prop}

Recall that $D_i \subseteq \ch_1$. Since $p_j \in \ch_2$, the segment $\overline{p_i p_j}$ must intersect $\gamma = \overline{q_{i_1} p_{i_1}} \cup \overline{p_{i_1} p_{i_2}} \cup \overline{p_{i_2} q_{i_2}}$. Hence, $\overline{p_i p_j}$ intersects one of the three segments $\overline{q_{i_1} p_{i_1}}$, $\overline{p_{i_1} p_{i_2}}$, or $\overline{p_{i_2} q_{i_2}}$.

If $\overline{p_i p_j}$ intersects $\overline{q_{i_1} p_{i_1}}$, then we have
\[
|p_i p_j| + r_{i_1} \stackrel{(1)}{=} |p_i p_j| + |p_{i_1} q_{i_1}| \stackrel{(2)}{\ge} |p_i p_{i_1}| + |p_j q_{i_1}| \stackrel{(3)}{>} r_{i_1} + r_i + r_j,
\]
where (1) follows from $|p_{i_1} q_{i_1}| = r_{i_1}$, (2) follows from the triangle inequality since $\overline{p_i p_j}$ intersects $\overline{q_{i_1} p_{i_1}}$, and (3) follows from $q_{i_1} \notin D_j$ and $D_i \cap D_{i_1} = \emptyset$. Therefore, $|p_i p_j| > r_i + r_j$, and thus $D_i$ and $D_j$ are disjoint, contradicting our assumption that they intersect. Hence, $\overline{p_i p_j}$ cannot intersect $\overline{q_{i_1} p_{i_1}}$.

By a similar argument, $\overline{p_i p_j}$ cannot intersect $\overline{q_{i_2} p_{i_2}}$. In what follows, we assume that $\overline{p_i p_j}$ only intersects $\overline{p_{i_1} p_{i_2}}$. Since $p_i \in \ch_1$, the point $p_j$ must lie below $\ell^*$. We have the following observation.

\begin{observation}\label{obser:contain}
$D_{i_1} \cap D_{i_2} \subseteq D_j$.
\end{observation}

\begin{proof}
Consider a point $x \in D_{i_1} \cap D_{i_2}$ that lies above $\ell^*$ (see Figure~\ref{fig:case3-50}). Note that the triangle $\triangle x p_{i_1} p_{i_2} \subseteq D_{i_1} \cup D_{i_2}$. Since $D_i$ is disjoint from $D_{i_1} \cup D_{i_2}$, we have $p_i \notin \triangle x p_{i_1} p_{i_2}$. Since $p_j$ lies below $\ell^*$, we also have $p_j \notin \triangle x p_{i_1} p_{i_2}$. Because $\overline{p_i p_j}$ intersects $\overline{p_{i_1} p_{i_2}}$, it must intersect either $\overline{x p_{i_1}}$ or $\overline{x p_{i_2}}$. We assume the former case; the latter can be handled analogously. We obtain
\[
r_{i_1} + r_j + r_i \stackrel{(1)}{\ge} |x p_{i_1}| + |p_i p_j| \stackrel{(2)}{\ge} |p_{i_1} p_i| + |x p_j| \stackrel{(3)}{>} r_{i_1} + r_i + |x p_j|,
\]
where (1) follows from $x \in D_{i_1}$ and $D_i \cap D_j \neq \emptyset$, (2) follows from the triangle inequality since $\overline{p_i p_j}$ intersects $\overline{x p_{i_1}}$, and (3) follows from $D_i \cap D_{i_1} = \emptyset$. Hence, $r_j > |x p_j|$, and thus $x \in D_j$.

\begin{figure}[t]
\begin{minipage}[h]{0.49\textwidth}
\begin{center}
\includegraphics[height=3.0in]{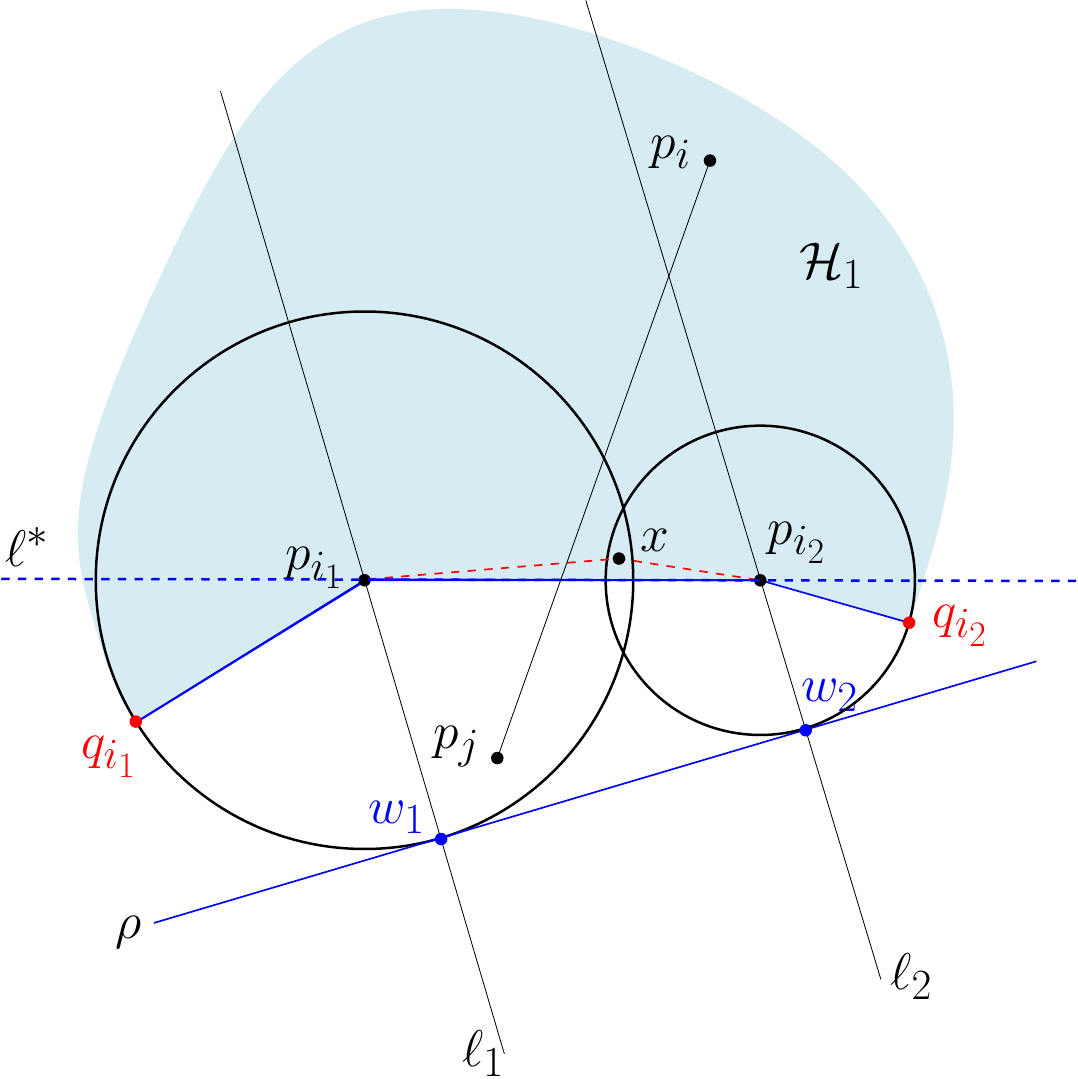}
\caption{\footnotesize Illustrating the proof of Observation~\ref{obser:contain}.}
\label{fig:case3-50}
\end{center}
\end{minipage}
\hspace{0.05in}
\begin{minipage}[h]{0.49\textwidth}
\begin{center}
\includegraphics[height=3.0in]{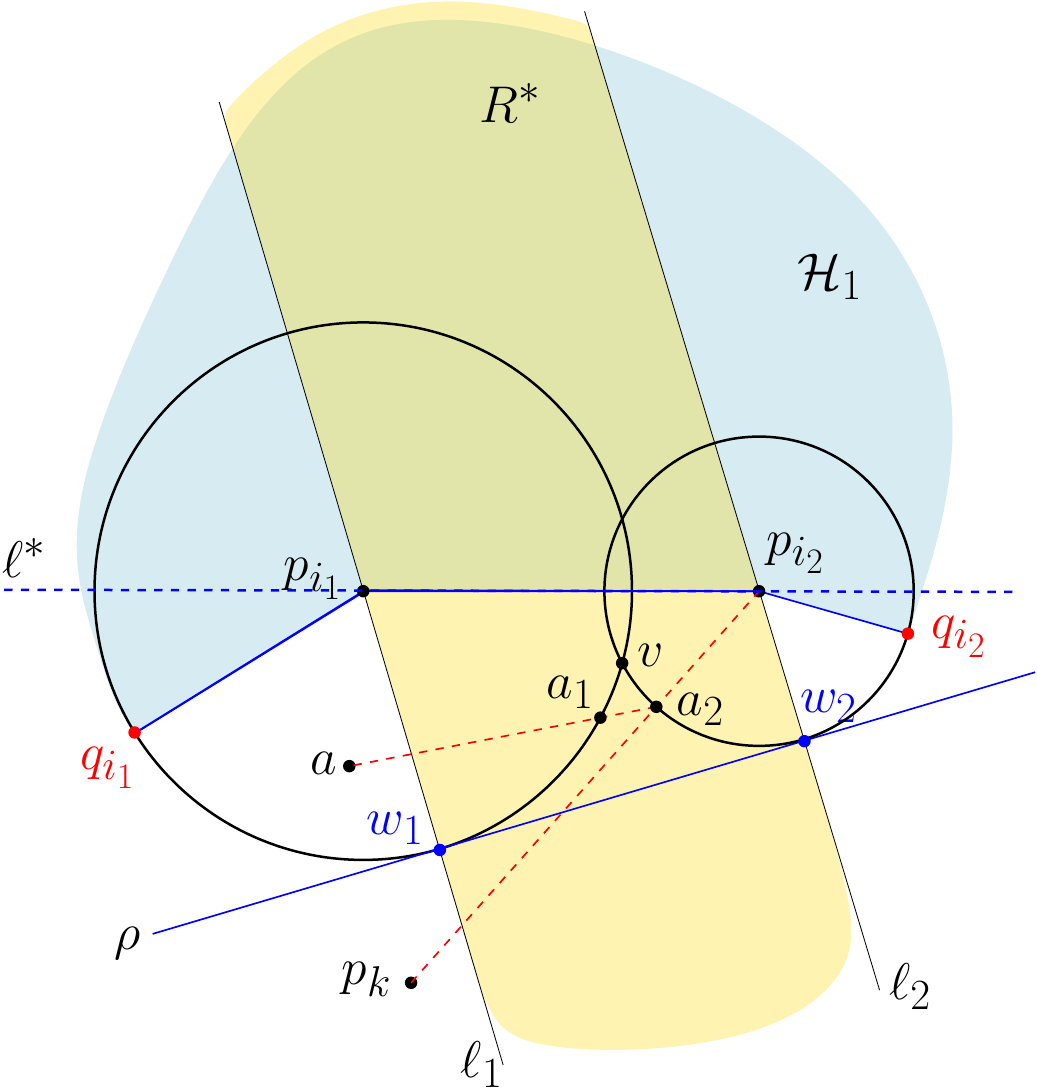}
\caption{\footnotesize Illustrating the proof of Observation~\ref{obser:40}.}
\label{fig:case3-60}
\end{center}
\end{minipage}
\vspace{-0.1in}
\end{figure}

This shows that all points of $D_{i_1} \cap D_{i_2}$ that lie above $\ell^*$ are contained in $D_j$. Now consider a point $x' \in D_{i_1} \cap D_{i_2}$ that lies below $\ell^*$. Let $x$ be the reflection of $x'$ with respect to $\ell^*$. Then $\ell^*$ is the perpendicular bisector of $x$ and $x'$, and every point below $\ell^*$ is closer to $x'$ than to $x$. Since $p_j$ lies below $\ell^*$, we have $|p_j x'| \le |p_j x| < r_j$. Hence, $x' \in D_j$.

This proves that $D_{i_1} \cap D_{i_2} \subseteq D_j$.
\end{proof}

In light of the above observation, if $D_k$ contains a point of $D_{i_1} \cap D_{i_2}$, then $D_k$ intersects $D_j$, and thus Proposition~\ref{prop:10} follows. In the following, we consider the case where $D_k$ does not intersect $D_{i_1} \cap D_{i_2}$. 

Recall that $p_k \in \ch_2$. If $p_k$ is above $\ell^*$, then by Lemma~\ref{lem:60}, $D_k$ must contain a point of $D_{i_1} \cap D_{i_2}$. Hence, this case cannot occur. In the following, we assume that $p_k$ is below $\ell^*$. 

Let $v$ be the intersection point of $\partial D_{i_1}$ and $\partial D_{i_2}$ that lies below $\ell^*$ (see Figure~\ref{fig:case3-60}). We have the following observation. 

\begin{observation}\label{obser:40}
$D_k$ contains a point $a_1$ on $\partial_{[w_1, v]} D_{i_1}$ and a point $a_2$ on $\partial_{[v, w_2]} D_{i_2}$.
\end{observation}

\begin{proof}
We distinguish two cases depending on whether $p_k \in R^*$. 

If $p_k \in R^*$, then since $p_k$ is below $\ell^*$ and $D_k$ intersects $D_{i_1}$ (resp., $D_{i_2}$), it follows that $D_k$ contains a point on $\partial_{[w_1, v]} D_{i_1}$ (resp., $\partial_{[v, w_2]} D_{i_2}$).

If $p_k \notin R^*$, then $p_k$ lies either to the left of $\ell_1$ or to the right of $\ell_2$. We assume the former case, as the latter can be treated analogously. 

We first claim that $p_k \notin D_{i_2}$. Indeed, suppose for contradiction that $p_k \in D_{i_2}$. Since $p_k$ is to the left of $\ell_1$, by the definition of $\ell_1$, $p_k$ must also lie in $D_{i_1}$. Hence, $p_k \in D_{i_1} \cap D_{i_2}$, and thus $D_k$ intersects $D_{i_1} \cap D_{i_2}$, which contradicts our assumption. The claim follows. 

Since $p_k \notin D_{i_2}$, the segment $\overline{p_k p_{i_2}}$ intersects $\partial D_{i_2}$ at a point $a_2$. Since $D_k$ intersects $D_{i_2}$, we have $a_2 \in D_k$. Hence, $a_2$ cannot lie in $D_{i_1} \cap D_{i_2}$, as otherwise $D_k$ would intersect $D_{i_1} \cap D_{i_2}$, contradicting our assumption. Consequently, since $p_k$ is below $\ell^*$, the point $a_2$ must lie on $\partial_{[v, w_2]} D_{i_2}$ (see Figure~\ref{fig:case3-60}).

Since $D_k$ intersects $D_{i_1}$ and $p_k$ is below $\ell^*$, the set $D_k \cap D_{i_1}$ contains a point $a$ below $\ell^*$. As both $a$ and $a_2$ lie in $D_k$, the segment $\overline{a a_2}$ is contained in $D_k$. Hence, $\overline{a a_2}$ does not intersect $D_{i_1} \cap D_{i_2}$, as otherwise $D_k$ would intersect $D_{i_1} \cap D_{i_2}$, contradicting our assumption. Therefore, $\overline{a a_2}$ must intersect the arc $\partial_{[w_1, v]} D_{i_1}$ at a point $a_1$. Since $\overline{a a_2} \subseteq D_k$ and $a_1\in \ol{aa_2}$, we have $a_1 \in D_k$.

This completes the proof.
\end{proof}

Let $x$ be a point in $D_{i_1} \cap D_{i_2}$. Since $p_j \in P(i_1,i_2)$, by the definitions of $q_{i_1}$, $q_j$, and $q_{i_2}$, the segment $\overline{x q_j}$ must intersect one of the following three portions: $\partial_{[q_{i_1}, w_1]} D_{i_1}$, $\overline{w_1 w_2}$, or $\partial_{[w_2, q_{i_2}]} D_{i_2}$. Since $a_1 \in \partial_{[w_1, v]} D_{i_1}$ and $a_2 \in \partial_{[v, w_2]} D_{i_2}$, it follows that $\overline{x q_j}$ must intersect one of the following: $\partial_{[q_{i_1}, a_1]} D_{i_1}$, $\overline{a_1 a_2}$, or $\partial_{[a_2, q_{i_2}]} D_{i_2}$. We argue below that in each case $D_k$ must intersect $D_j$.

Note that since both $x$ and $q_j$ lie in $D_j$, we have $\overline{x q_j} \subseteq D_j$. Similarly, since both $a_1$ and $a_2$ lie in $D_k$, we have $\overline{a_1 a_2} \subseteq D_k$. 

\begin{itemize}
    \item If $\overline{x q_j}$ intersects $\overline{a_1 a_2}$, then since $\overline{x q_j} \subseteq D_j$ and $\overline{a_1 a_2} \subseteq D_k$, it follows that $D_j$ and $D_k$ intersect. 

    \item If $\overline{x q_j}$ intersects $\partial_{[q_{i_1}, a_1]} D_{i_1}$, let $q$ be the intersection point (see Figure~\ref{fig:case3-70}). Then $q_{i_1}$, $q$, $a_1$, and $v$ appear in this counterclockwise order along $\partial D_{i_1}$. Since $\overline{x q_j} \subseteq D_j$ and $q\in \overline{xq_j}$, we have $q \in D_j$. 

\begin{figure}[t]
\begin{minipage}[h]{\textwidth}
\begin{center}
\includegraphics[height=3.0in]{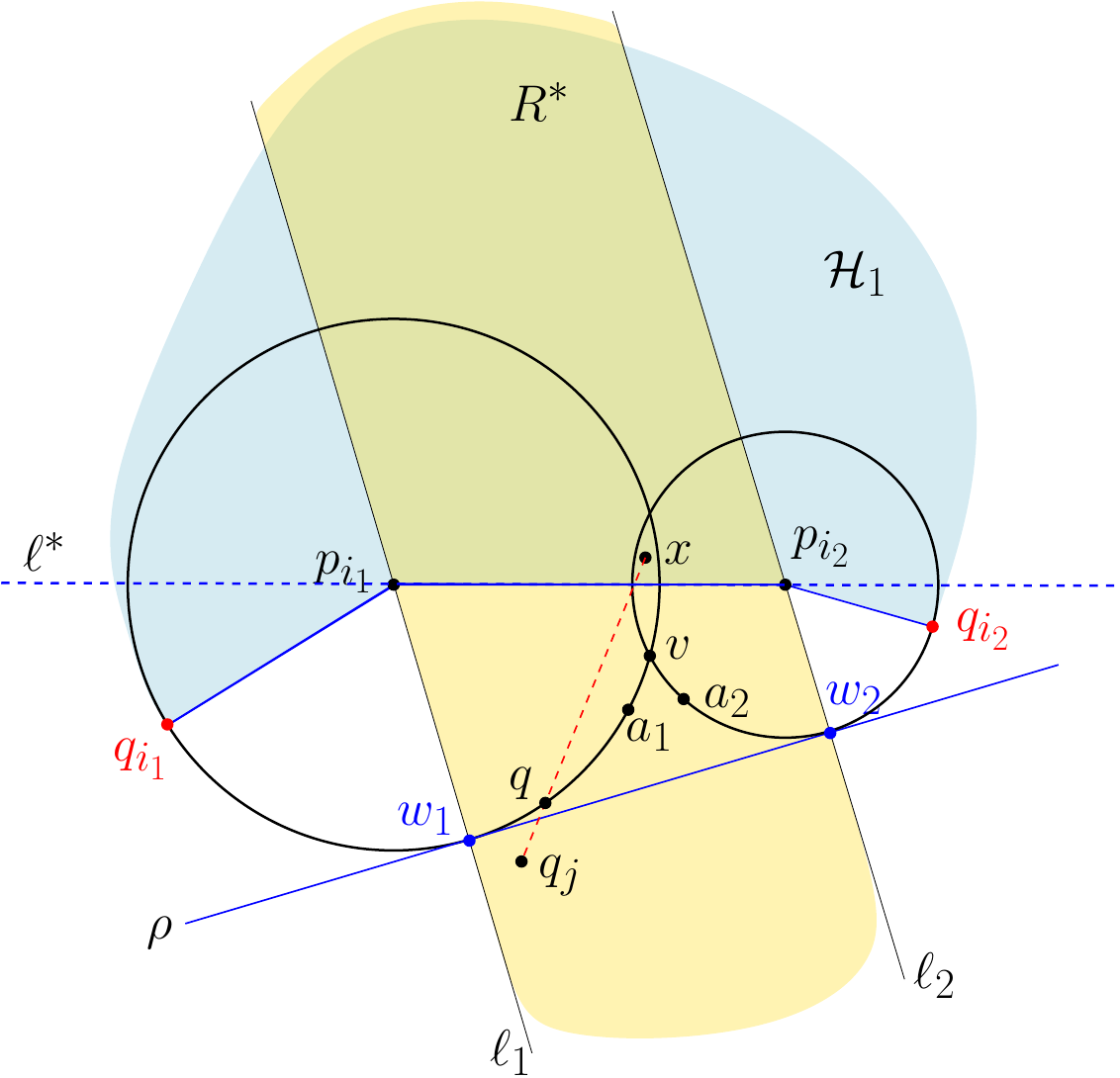}
\caption{\footnotesize Illustrating the proof of Proposition~\ref{prop:10}.}
\label{fig:case3-70}
\end{center}
\end{minipage}
\vspace{-0.1in}
\end{figure}

    We claim that $a_1 \in D_j$. Suppose, for contradiction, that $a_1 \notin D_j$. By Observation~\ref{obser:contain}, we have $v \in D_j$. Hence, $D_j$ contains $q$ and $v$, but does not contain $q_{i_1}$ or $a_1$. It follows that $\partial D_j$ must intersect $\partial D_{i_1}$ at least three times: once on $\partial_{[q_{i_1}, q]} D_{i_1}$, once on $\partial_{[q, a_1]} D_{i_1}$, and once on $\partial_{[a_1, v]} D_{i_1}$. However, this is impossible, since two circles can intersect in at most two points. Therefore, $a_1 \in D_j$.

    Since $a_1 \in D_j$ and $a_1 \in D_k$ by Observation~\ref{obser:40}, it follows that $D_j$ intersects $D_k$. 

    \item If $\overline{x q_j}$ intersects $\partial_{[a_2, q_{i_2}]} D_{i_2}$, then by an argument symmetric to the previous case, we also obtain that $D_j$ intersects $D_k$. 
\end{itemize}

In summary, $D_j$ intersects $D_k$ in all cases, and thus Proposition~\ref{prop:10} holds. 

This completes the proof of Case~3 and, consequently, the proof of Lemma~\ref{lem:key}.

\section{The general case}
\label{sec:general}

In this section, we consider the general case, in which the disks of $\cd$ are not necessarily in any convex position. Our goal is to compute an independent set of $\cd$ of size $3$, if one exists.

In the following, we first present a simple $O(n^2 \log n)$-time algorithm, and then improve it to a subquadratic algorithm. We follow the notation introduced in Section~\ref{sec:pre}. 

\subsection{An $\boldsymbol{O(n^2 \log n)}$-time algorithm}

For each disk $D_j \in \cd$, define $\cd_j = \{ D \in \cd \mid D \cap D_j = \emptyset \}$; that is, $\cd_j$ consists of all disks in $\cd$ that are disjoint from $D_j$.

For each $j \in [1,n]$, we construct the farthest-disk Voronoi diagram $\mathcal{FVD}(\cd_j)$ for the disks in $\cd_j$, together with a point-location data structure on $\mathcal{FVD}(\cd_j)$. Constructing $\mathcal{FVD}(\cd_j)$ takes $O(|\cd_j| \log |\cd_j|)$ time~\cite{ref:RappaportCo89}, and the point-location structure can be built in an additional $O(|\cd_j|)$ time~\cite{ref:EdelsbrunnerOp86,ref:KirkpatrickOp83}. 

Since $|\cd_j| \le n$ for every $j$, constructing these structures for all $j \in [1,n]$ takes
\[
\sum_{j=1}^{n} O(|\cd_j| \log |\cd_j|) = O(n^2 \log n)
\]
time in total. Using $\mathcal{FVD}(\cd_j)$, given a point $q$, the farthest disk of $\cd_j$ from $q$ can be found in $O(\log n)$ time. Note that the \emph{distance} from a point $q$ to a disk $D_i$ is defined as $|q p_i| - r_i$.

For each pair $(p_i, p_j)$ such that $D_i$ and $D_j$ are disjoint, we query the center $p_i$ in $\mathcal{FVD}(\cd_j)$ and find the disk $D_k \in \cd_j$ that is farthest from $p_i$ in $O(\log n)$ time. If $D_i$ and $D_k$ are disjoint, then $\{D_i, D_j, D_k\}$ is an independent set. Otherwise, no disk in $\cd_j$ is disjoint from $D_i$, and thus $\cd$ does not contain an independent set of size $3$ that includes the pair $\{D_i, D_j\}$.

In this way, an independent set of size $3$ can be found, if it exists, in a total of $O(n^2 \log n)$ time.

\subsection{A sub-quadratic algorithm}

The bottleneck in the above algorithm is that the sets $\cd_j$ are handled independently, even though they may overlap heavily. Moreover, it inspects every pair $(p_i, p_j)$, which already costs $\Omega(n^2)$ time. To avoid this repeated work, we resort to a data structure for semi-algebraic range searching~\cite{ref:AgarwalIn25}. 

Let $\cp$ denote the set of ordered pairs $(p_i, p_j)$ of $P$ such that their corresponding disks $D_i$ and $D_j$ are disjoint. We have the following definition of \emph{biclique partitions}. 

\begin{definition}
\label{def:biclique-general}
A \emph{biclique partition} of $\cp$ is a collection $\Gamma(\cp)=\{A_t \times B_t \mid A_t, B_t \subseteq P\}$
such that:
\begin{enumerate}
    \item for every pair $(p_i, p_j) \in A_t \times B_t$, the disks $D_i$ and $D_j$ are disjoint;
    \item for every pair $(p_i, p_j)$ such that $D_i$ and $D_j$ are disjoint, there is a unique biclique $A_t \times B_t \in \Gamma(\cp)$ that contains $(p_i, p_j)$.
\end{enumerate}
\end{definition}

Moreover, our algorithm requires a stronger version of a biclique partition, called a \emph{tree-structured biclique partition}, defined below. This concept was originally introduced in~\cite{ref:TkachenkoDo25}. 

\begin{definition}[Tree-structured biclique partition]
\label{def:tree-biclique}
A biclique partition $\Gamma(\cp) = \{A_t \times B_t \mid A_t, B_t \subseteq P\}$ is \emph{tree-structured} if all the subsets $A_t$ form a tree $\ct_A$ such that, for each internal node $A_t$, its children subsets form a partition of $A_t$. 
\end{definition}

Note that in the tree $\ct_A$ in Definition~\ref{def:tree-biclique}, we consider a node to be an \emph{ancestor} of itself. By definition, if $A_{t'}$ is an ancestor of $A_t$ in $\ct_A$, then $A_t \subseteq A_{t'}$. 

The next lemma, analogous to one presented in~\cite{ref:TkachenkoDo25}, explains why a tree-structured biclique partition is useful for finding an independent set of size $3$. 

\begin{lemma}
\label{lem:treecondition}
Suppose that $\Gamma(\cp)=\{A_t \times B_t \mid A_t, B_t \subseteq P\}$
is a tree-structured biclique partition of $\cp$, and let $\mathcal{T}_A$ be the tree formed by the subsets $A_t$. Then, three disks $D_i, D_j, D_k \in \cd$ form an independent set if and only if there exists a biclique $A_t \times B_t \in \Gamma(\cp)$ containing $(p_i, p_j)$ such that $A_t$ has an ancestor subset $A_{t'}$ in $\mathcal{T}_A$ with $p_k \in B_{t'}$ and $D_j \cap D_k = \emptyset$. 
\end{lemma}

\begin{proof}
Assume that $\{D_i, D_j, D_k\}$ is an independent set. Since $D_i$ and $D_j$ are disjoint, by Definition~\ref{def:biclique-general}, there is a unique biclique $A_t \times B_t$ containing $(p_i, p_j)$. Similarly, since $D_i$ and $D_k$ are disjoint, there is a unique biclique $A_{t'} \times B_{t'}$ containing $(p_i, p_k)$.
As both $A_t$ and $A_{t'}$ contain $p_i$ and $\Gamma(\cp)$ is tree-structured, one of $A_t$ and $A_{t'}$ must be an ancestor of the other. If $A_{t'}$ is an ancestor of $A_t$, then the required condition in the lemma holds. Otherwise, we simply exchange the roles of $D_j$ and $D_k$.

Conversely, suppose that $\Gamma(\cp)$ has a biclique $A_t \times B_t$ containing $(p_i, p_j)$ and that $A_t$ has an ancestor $A_{t'}$ such that $p_k \in B_{t'}$ and $D_j \cap D_k = \emptyset$. Since $(p_i, p_j) \in A_t \times B_t$, $D_i$ and $D_j$ are disjoint. Because $A_{t'}$ is an ancestor of $A_t$, we have $A_t \subseteq A_{t'}$, and thus $p_i \in A_{t'}$. Since $p_k \in B_{t'}$, it follows that $(p_i, p_k) \in A_{t'} \times B_{t'}$, so $D_i$ and $D_k$ are disjoint. Therefore, $D_i$, $D_j$, and $D_k$ are pairwise disjoint and hence form an independent set. 
\end{proof}

Using a semi-algebraic range searching data structure given in~\cite{ref:AgarwalIn25}, the following lemma constructs a tree-structured biclique partition. 

\begin{lemma}\label{lem:algotree}
A tree-structured biclique partition $\Gamma(\cp) = \{A_t \times B_t \mid A_t, B_t \subseteq P\}$, together with the tree $\ct_A$, can be constructed in $O^*(n^{5/3})$ randomized expected time such that $\sum_t |A_t| = O^*(n)$, $\sum_t |B_t| = O^*(n^{5/3})$, and the height of $\ct_A$ is $O^*(1)$, where $O^*$ hides a factor of $n^{\epsilon}$ for an arbitrarily small $\epsilon > 0$. 
\end{lemma}

\begin{proof}
For each point $p_i \in P$, let $x_i$ and $y_i$ denote its $x$- and $y$-coordinates, respectively. Recall that $r_i$ is the radius of $D_i$. We ``lift'' $p_i$ to a point $q_i = (x_i, y_i, r_i)$ in $\bbR^3$, and define a semi-algebraic range $R_i$ in $\bbR^3$:
\[
R_i = \Bigl\{(x,y,z) \in \bbR^3 \ \Big| \ \sqrt{(x - x_i)^2 + (y - y_i)^2} > z + r_i \Bigr\}.
\]

For any two disks $D_i, D_j \in \cd$, observe that
\[
q_i \in R_j \Longleftrightarrow |p_i p_j| > r_i + r_j \Longleftrightarrow D_i \cap D_j = \emptyset.
\]

Define $Q = \{q_i \mid p_i \in P\}$ and $\calR = \{R_i \mid p_i \in P\}$. Following the notation in~\cite{ref:AgarwalIn25}, we define a predicate $\Pi$ on $Q \times \calR$ by $\Pi(q_i, R_j) = 1 \Longleftrightarrow q_i \in R_j$. Both families $Q$ and $\calR$ have three degrees of freedom, and $\Pi$ is a constant-complexity semi-algebraic predicate defined by a single polynomial inequality (i.e., $k = 1$ in the notation of~\cite[Section~A.1]{ref:AgarwalIn25}).

We construct a semi-algebraic range searching data structure $\Psi$ for $Q$ with respect to the ranges in $\calR$ using~\cite[Theorem~A.3]{ref:AgarwalIn25}. The data structure $\Psi$ is a tree in which each node $t \in \Psi$ is associated with a \emph{canonical subset} $Q_t$ of $Q$. The canonical subsets $Q_t$ satisfy the tree-structured property required by Definition~\ref{def:tree-biclique}: for each internal node $t' \in \Psi$, the subsets $Q_t$ corresponding to the children $t$ of $t'$ form a partition of $Q_{t'}$~\cite[Section~A.1]{ref:AgarwalIn25}. 

To construct a tree-structured biclique partition $\Gamma(\cp) = \{A_t \times B_t \mid A_t, B_t \subseteq P\}$, for each node $t \in \Psi$, we let $A_t = \{p_i \mid q_i \in Q_t\}$, and let $\ct_A$ be the same tree structure as $\Psi$. 

The data structure $\Psi$ can be constructed in randomized expected $O^*(n)$ time, and the total size of all canonical subsets is $O^*(n)$~\cite[Section~A.1]{ref:AgarwalIn25}. Hence, $\sum_t |A_t| = O^*(n)$. In addition, the height of $\Psi$ is $O^*(1)$~\cite[Section~A.1]{ref:AgarwalIn25}, and thus the height of $\ct_A$ is also $O^*(1)$. 

We construct the subsets $B_t$ as follows. For each range $R_j \in \calR$, we perform a range query on $\Psi$. The query algorithm returns, in $O^*(n^{2/3})$ time, a subset $V_{R_j}$ of $O^*(n^{2/3})$ nodes $t$ whose canonical subsets $Q_t$ form a partition of $R_j \cap Q$~\cite[Section~A.1]{ref:AgarwalIn25}. For each node $t \in V_{R_j}$, we add $p_j$ to $B_t$. 

In this way, we obtain a collection of subsets $B_t$ whose total size is
\[
\sum_{R_j \in \calR} |V_{R_j}| = O^*(n^{5/3}),
\]
and all $A_t \times B_t$ thus obtained form a tree-structured biclique partition of $\cp$. The total time for constructing all subsets $B_t$ is $O^*(n^{5/3})$.

In summary, the overall time for computing $\Gamma(\cp)$ is $O^*(n)$ randomized expected time plus $O^*(n^{5/3})$ deterministic time.
\end{proof}

\paragraph{The algorithm: Computing an independent set of size $3$.}
With the tree-structured biclique partition $\Gamma(\cp)$ computed in Lemma~\ref{lem:algotree}, we now present our algorithm for computing an independent set of size $3$. Our algorithm is based on Lemma~\ref{lem:treecondition}. 

We first perform certain preprocessing work for the subsets $B_t$. 
For each subset $B_t$ of $\Gamma(\cp)$, we construct the farthest-disk Voronoi diagram $\mathcal{FVD}(B_t)$ for the disks in $B_t$~\cite{ref:RappaportCo89}, together with a point-location data structure on $\mathcal{FVD}(B_t)$~\cite{ref:EdelsbrunnerOp86,ref:KirkpatrickOp83}. Constructing $\mathcal{FVD}(B_t)$ takes $O(|B_t|\log |B_t|)$ time, and constructing the point-location structure takes an additional $O(|B_t|)$ time. 
Since $\sum_t |B_t| = O^*(n^{5/3})$ by Lemma~\ref{lem:algotree}, performing this preprocessing for all subsets $B_t$ takes $O^*(n^{5/3})$ time in total. 

We next search for an independent set of size $3$. 

For each subset $B_t$ of $\Gamma(\cp)$, for each point $p_j \in B_t$, and for every ancestor $A_{t'}$ of $A_t$, we find the farthest disk $D_k$ in $B_{t'}$ from $p_j$. This can be done in $O(\log n)$ time using a point-location query on $\mathcal{FVD}(B_{t'})$. 
If $D_j$ and $D_k$ are disjoint, then by Lemma~\ref{lem:treecondition}, $\{D_i, D_j, D_k\}$ is an independent set for any point $p_i \in A_t$, and we can report such a set and terminate the algorithm. 

If the above procedure fails to report any independent set, then by Lemma~\ref{lem:treecondition} no independent set of size $3$ exists in $\cd$. 

Since the height of $\ct_A$ is $O^*(1)$ and $\sum_t |B_t| = O^*(n^{5/3})$ by Lemma~\ref{lem:algotree}, the total time of the above searching procedure is $O^*(n^{5/3})$.

Therefore, the overall time of the algorithm is $O^*(n^{5/3})$. We conclude this section with the following theorem.

\begin{theorem}
\label{thm:general-case}
Given a set $\cd$ of $n$ disks in the plane, one can compute, in $O^*(n^{5/3})$ randomized expected time, a subset of three pairwise disjoint disks, or report that no such subset exists.
\end{theorem}





\bibliographystyle{plainurl}
\bibliography{refs}

\end{document}